\documentclass[a4paper,11pt]{article}
\usepackage{jcappub} % for details on the use of the package, please see the JINST-author-manual
\usepackage{lineno}
\usepackage{graphicx}	% Including figure files
\usepackage{subcaption}
\usepackage{grffile}

\usepackage{amsmath}	% Advanced maths commands
\usepackage{mathtools}
\usepackage{physics}
\usepackage{xcolor}

\def\setsymbol#1#2{\expandafter\def\csname #1\endcsname{#2}}
\def\getsymbol#1{\csname #1\endcsname}

\newbox\tablebox    \newdimen\tablewidth
\def\leaderfil{\leaders\hbox to 5pt{\hss.\hss}\hfil}
\def\tablenote#1 #2\par{\begingroup \parindent=0.8em
    \abovedisplayshortskip=0pt\belowdisplayshortskip=0pt
    \noindent
    $$\hss\vbox{\hsize\tablewidth \hangindent=\parindent \hangafter=1 \noindent
    \hbox to \parindent{$^#1$\hss}\strut#2\strut\par}\hss$$
    \endgroup}

\def\L2{\ifmmode L_2\else $L_2$\fi}

\def\DeltaT{\ifmmode \Delta T\else $\Delta T$\fi}
\def\deltat{\ifmmode \Delta t\else $\Delta t$\fi}
\def\fknee{\ifmmode f_{\rm knee}\else $f_{\rm knee}$\fi}
\def\Fmax{\ifmmode F_{\rm max}\else $F_{\rm max}$\fi}
\def\solar{\ifmmode{\rm M}_{\mathord\odot}\else${\rm M}_{\mathord\odot}$\fi}
\def\Msolar{\ifmmode{\rm M}_{\mathord\odot}\else${\rm M}_{\mathord\odot}$\fi}
\def\Lsolar{\ifmmode{\rm L}_{\mathord\odot}\else${\rm L}_{\mathord\odot}$\fi}
\def\inv{\ifmmode^{-1}\else$^{-1}$\fi}
\def\mo{\ifmmode^{-1}\else$^{-1}$\fi}
\def\sup#1{\ifmmode ^{\rm #1}\else $^{\rm #1}$\fi}
\def\expo#1{\ifmmode \times 10^{#1}\else $\times 10^{#1}$\fi}
\def\,{\thinspace}
\def\lsim{\mathrel{\raise .4ex\hbox{\rlap{$<$}\lower 1.2ex\hbox{$\sim$}}}}
\def\gsim{\mathrel{\raise .4ex\hbox{\rlap{$>$}\lower 1.2ex\hbox{$\sim$}}}}

\def\simprop{\mathrel{\raise .4ex\hbox{\rlap{$\propto$}\lower 1.2ex\hbox{$\sim$}}}}
\def\deg{\ifmmode^\circ\else$^\circ$\fi}
\def\pdeg{\ifmmode $\setbox0=\hbox{$^{\circ}$}\rlap{\hskip.11\wd0 .}$^{\circ}
          \else \setbox0=\hbox{$^{\circ}$}\rlap{\hskip.11\wd0 .}$^{\circ}$\fi}
\def\arcs{\ifmmode {^{\scriptstyle\prime\prime}}
          \else $^{\scriptstyle\prime\prime}$\fi}
\def\arcm{\ifmmode {^{\scriptstyle\prime}}
          \else $^{\scriptstyle\prime}$\fi}
\newdimen\sa  \newdimen\sb
\def\parcs{\sa=.07em \sb=.03em
     \ifmmode \hbox{\rlap{.}}^{\scriptstyle\prime\kern -\sb\prime}\hbox{\kern -\sa}
     \else \rlap{.}$^{\scriptstyle\prime\kern -\sb\prime}$\kern -\sa\fi}
\def\parcm{\sa=.08em \sb=.03em
     \ifmmode \hbox{\rlap{.}\kern\sa}^{\scriptstyle\prime}\hbox{\kern-\sb}
     \else \rlap{.}\kern\sa$^{\scriptstyle\prime}$\kern-\sb\fi}
\def\ra[#1 #2 #3.#4]{#1\sup{h}#2\sup{m}#3\sup{s}\llap.#4}
\def\dec[#1 #2 #3.#4]{#1\deg#2\arcm#3\arcs\llap.#4}
\def\deco[#1 #2 #3]{#1\deg#2\arcm#3\arcs}
\def\rra[#1 #2]{#1\sup{h}#2\sup{m}}

\def\dots{\relax\ifmmode \ldots\else $\ldots$\fi}
\def\WHzsr{\ifmmode $W\,Hz\mo\,sr\mo$\else W\,Hz\mo\,sr\mo\fi}
\def\mHz{\ifmmode $\,mHz$\else \,mHz\fi}
\def\GHz{\ifmmode $\,GHz$\else \,GHz\fi}
\def\mKs{\ifmmode $\,mK\,s$^{1/2}\else \,mK\,s$^{1/2}$\fi}
\def\muKs{\ifmmode \,\mu$K\,s$^{1/2}\else \,$\mu$K\,s$^{1/2}$\fi}
\def\muKRJs{\ifmmode \,\mu$K$_{\rm RJ}$\,s$^{1/2}\else \,$\mu$K$_{\rm RJ}$\,s$^{1/2}$\fi}
\def\muKHz{\ifmmode \,\mu$K\,Hz$^{-1/2}\else \,$\mu$K\,Hz$^{-1/2}$\fi}
\def\MJysr{\ifmmode \,$MJy\,sr\mo$\else \,MJy\,sr\mo\fi}
\def\MJysrmK{\ifmmode \,$MJy\,sr\mo$\,mK$_{\rm CMB}\mo\else \,MJy\,sr\mo\,mK$_{\rm CMB}\mo$\fi}
\def\microns{\ifmmode \,\mu$m$\else \,$\mu$m\fi}
\def\micron{\microns}
\def\muK{\ifmmode \,\mu$K$\else \,$\mu$\hbox{K}\fi}
\def\microK{\ifmmode \,\mu$K$\else \,$\mu$\hbox{K}\fi}
\def\muW{\ifmmode \,\mu$W$\else \,$\mu$\hbox{W}\fi}
\def\kms{\ifmmode $\,km\,s$^{-1}\else \,km\,s$^{-1}$\fi}
\def\kmsMpc{\ifmmode $\,\kms\,Mpc\mo$\else \,\kms\,Mpc\mo\fi}
\providecommand{\sorthelp}[1]{}

\usepackage{booktabs}
\usepackage{multirow}
\usepackage{makecell}
\usepackage{tabulary}
\usepackage{microtype}

\newcommand{\bde}{{\bf e }}
\newcommand{\bdx}{{\bf x }}

\newcommand{\bdw}{{\bf w }}

\newcommand{\bda}{{\bf a }}

\newcommand{\bdg}{{\bf g }}

\newcommand{\bdt}{{\bf t }}

\newcommand{\tA}{{\bf \rm A}}

\newcommand{\tC}{{\bf \rm C}}

\newcommand{\tI}{{\bf \rm I}}
\newcommand{\tS}{{\bf \rm S}}

\newcommand{\tN}{{\bf \rm N}}

\newcommand{\tP}{{\bf \rm P}}

\newcommand{\lb}{\emph{LiteBIRD}}
\newcommand{\pl}{\emph{Planck}}

\newlength{\arrow}
\def\creff@jnl#1{{\rm#1\/}}

\def\aj{\creff@jnl{AJ}}                  % Astronomical Journal
\def\araa{\creff@jnl{ARA\&A}}            % Annual Review of Astron and Astrophys
\def\apj{\creff@jnl{ApJ}}                % Astrophysical Journal
\def\apjl{\creff@jnl{ApJ}}               % Astrophysical Journal, Letters
\def\apjs{\creff@jnl{ApJS}}              % Astrophysical Journal, Supplement
\def\ao{\creff@jnl{Appl.Optics}}         % Applied Optics
\def\apss{\creff@jnl{Ap\&SS}}            % Astrophysics and Space Science
\def\aap{\creff@jnl{A\&A}}               % Astronomy and Astrophysics
\def\aapr{\creff@jnl{A\&A~Rev.}}         % Astronomy and Astrophysics Reviews
\def\aaps{\creff@jnl{A\&AS}}             % Astronomy and Astrophysics, Supplement
\def\azh{\creff@jnl{AZh}}                % Astronomicheskii Zhurnal
\def\baas{\creff@jnl{BAAS}}              % Bulletin of the AAS
\def\jcap{\creff@jnl{JCAP}}              % Journal of Cosmology and Astroparticle Physics
\def\jrasc{\creff@jnl{JRASC}}            % Journal of the RAS of Canada
\def\memras{\creff@jnl{MmRAS}}           % Memoirs of the RAS
\def\mnras{\creff@jnl{MNRAS}}            % Monthly Notices of the RAS
\def\pra{\creff@jnl{Phys.Rev.A}}         % Physical Review A: General Physics
\def\prb{\creff@jnl{Phys.Rev.B}}         % Physical Review B: Solid State
\def\prc{\creff@jnl{Phys.Rev.C}}         % Physical Review C
\def\prd{\creff@jnl{Phys.Rev.D}}         % Physical Review D
\def\prl{\creff@jnl{Phys.Rev.Lett}}      % Physical Review Letters
\def\physrep{\creff@jnl{Phys.Rep.}}      % Physics Report
\def\pasp{\creff@jnl{PASP}}              % Publications of the ASP
\def\pasj{\creff@jnl{PASJ}}              % Publications of the ASJ
\def\qjras{\creff@jnl{QJRAS}}            % Quarterly Journal of the RAS
\def\skytel{\creff@jnl{S\&T}}            % Sky and Telescope
\def\solphys{\creff@jnl{Solar~Phys.}}    % Solar Physics
\def\sovast{\creff@jnl{Soviet~Ast.}}     % Soviet Astronomy
\def\ssr{\creff@jnl{Space~Sci.Rev.}}     % Space Science Reviews
\def\zap{\creff@jnl{ZAp}}                % Zeitschrift fuer Astrophysik
\def\nat{\creff@jnl{Nature}}             % Nature 

\title{\boldmath Enhanced foreground mitigation in thermal SZ Compton-$y$ maps via polarization and deprojection}

\author{Jyothis Chandran,}
\author{Mathieu Remazeilles,}
\author{R.~B. Barreiro}
\affiliation{Instituto de Física de Cantabria (CSIC-UC),\\
Avda. de los Castros s/n, 39005 Santander, Spain}

\emailAdd{chandran@ifca.unican.es}

\abstract{Residual foreground contamination in thermal Sunyaev-Zeldovich (SZ) Compton-$y$ parameter maps ($y$-maps) arises mainly from Galactic emissions---thermal dust and synchrotron radiation---on large angular scales, and from cosmic infrared background (CIB) anisotropies on small scales. Unlike the thermal SZ effect, Galactic foregrounds are strongly polarized. Exploiting this distinction, we introduce a \emph{hybrid} Needlet Internal Linear Combination (Hybrid NILC) method that combines \pl\ total-intensity and polarization frequency maps in the component-separation pipeline, thereby improving the suppression of residual Galactic emission while preserving the unpolarized SZ signal by leveraging the intrinsic $TE$ and $TB$ correlations of thermal dust and synchrotron. Using \pl\ PR4 data, we find that the Hybrid NILC $y$-map exhibits about $40\,\%$ lower cross-correlation with the IRAS dust tracer than the standard temperature-only \pl\ $y$-map, indicating reduced residual Galactic contamination. Simulations further indicate that, for future high-sensitivity surveys such as \lb, the Hybrid NILC will become increasingly effective at suppressing Galactic residuals. We further address small-scale extragalactic contamination by selectively deprojecting specific moments of the CIB using a Constrained Hybrid NILC variant, achieving an improved balance between CIB suppression and noise penalty compared to previous implementations in the literature. These novel approaches---particularly the joint use of temperature and polarization in component separation---offer a powerful framework for disentangling polarized and unpolarized signals.}


\begin{document}

\maketitle
\flushbottom

%%%%%%%%%%%%%%%%%%%%%%%%%%%%%%%%%%%%%%%% BODY OF PAPER %%%%%%%%%%%%%%%%%%%%%%%%%%%%%%%%%%%%%%%%%%%%%

\section{Introduction}
\label{sec:intro}

As cosmic microwave background (CMB) photons travel from the last-scattering surface to the observer, they pass through hot, ionized gas within and between galaxy clusters, where they are upscattered to higher energies via inverse Compton scattering with energetic free electrons. This process, known as the thermal Sunyaev-Zeldovich (SZ) effect \citep{1969SunyaevZeldovich, 1972SunyaevZeldovich}, induces anisotropic $y$-type spectral distortions of the CMB blackbody radiation that trace the distribution of hot gas. The thermal SZ $y$-distortion is independent of redshift and proportional to the electron gas pressure integrated along the line of sight, providing a unique and powerful probe of the hot baryonic component of the Universe. Its distinctive spectral signature enables its extraction through multi-frequency submillimetre sky observations and component-separation techniques \citep{2002MNRAS.336.1057H, 2002MNRAS.336.1351D, 2008A&A...491..597L, 2011MNRAS.410.2481R, 2013MNRAS.430..370R, 2013A&A...558A.118H}, allowing the hot gas distribution—and hence the large-scale structure (LSS) of the Universe, including both galaxy clusters and the diffuse, unbound filamentary gas connecting them—to be mapped across the entire sky.

The extraction of the thermal SZ (tSZ) signal from multi-frequency sky maps has enabled the construction of several galaxy cluster catalogues over the past decade \citep{planck2014-a36, SPT-SZ-Cat2015, ACT-SZ-Cat2021, Melin2021, ACT-SZ-Cat2026}, as well as Compton $y$-parameter maps ($y$-maps) that trace the diffuse hot gas distribution beyond the well-resolved, massive clusters \citep{planck2014-a28, 2020PhRvD.102b3534M, 2022ApJS..258...36B, 2022MNRAS.509..300T, PR4NILCymap, NILC-McCarthyHill,PhysRevD.109.063530,Maniyar2026}. The statistics of cluster number counts and of the $y$-map, such as its power spectrum and bispectrum, as well as cross-correlations between the $y$-map and other LSS tracers, provide constraints on key cosmological parameters, including the total matter density ($\Omega_m$) and the amplitude of matter clustering on $8\,h^{-1}\text{Mpc}$ scales ($\sigma_8$) \citep{planck2013-p15, planck2013-p05b, planck2014-a30, planck2014-a28, 2016ApJ...832...95D, Bolliet2018, 2019ApJ...878...55B, 2022MNRAS.509..300T}, establishing the thermal SZ effect as a powerful cosmological probe that is complementary to,  and largely independent of, the primary CMB anisotropies \citep{1999PhR...310...97B,Komatsu1999,2002ARA&A..40..643C,Komatsu2002,Rubino-Martin2003,2019SSRv..215...17M}.

However, the full cosmological potential of the thermal SZ effect remains limited by residual contamination from instrumental noise and astrophysical foregrounds. For current datasets such as \pl, the Atacama Cosmology Telescope (ACT), and the South Pole Telescope (SPT), instrumental noise has been a major limiting factor (see, e.g., Appendix~A of \cite{PR4NILCymap}). In contrast, for upcoming CMB surveys with substantially higher sensitivity, such as \lb\ \citep{LB2022}, and improved angular resolution, such as the Simons Observatory (SO) \citep{SO-LAT2025}, residual foreground contamination is expected to become the primary source of uncertainty (see, e.g., \cite{LiteBIRD:2024dbi}). This motivates the development of novel foreground-mitigation strategies that go beyond temperature-only information to fully exploit future tSZ $y$-maps.

The main foreground contaminants in $y$-maps arise from Galactic emission at large angular scales (primarily thermal dust and synchrotron), and from extragalactic cosmic infrared background (CIB) anisotropies from dusty star-forming galaxies at small angular scales \cite{planck2014-a28, PR4NILCymap,LiteBIRD:2024dbi}. In previous work \cite{PR4NILCymap}, we produced and publicly released a new all-sky thermal SZ $y$-map with the lowest contamination variance among existing products, using a tailored Needlet Internal Linear Combination (NILC) component-separation pipeline \citep{2009A&A...493..835D} applied to the \pl\ Release 4 (PR4) data \cite{planck2020-LVII}. In the present study, we investigate extensions of this NILC framework aimed at further improving foreground mitigation in the \pl\ $y$-map and demonstrate the potential of these extensions for tSZ reconstruction in future high-sensitivity CMB surveys.

Galactic foregrounds such as thermal dust and synchrotron emissions primarily contaminate the tSZ signal at low Galactic latitudes and on large angular scales, thereby obscuring diffuse, large-scale SZ emission from two-halo clustering \citep{Komatsu1999} as well as contributions from the most massive, low redshift clusters \citep{Komatsu2002}. These Galactic emissions also extend to relatively high Galactic latitudes compared to free-free emission, making them increasingly relevant for future high-sensitivity SZ surveys. Their complex spatial variability in spectral energy distribution (SED) renders parametric modelling challenging, favoring blind component-separation methods with both spatial and harmonic localization, such as NILC (see Appendix~B of \cite{PR4NILCymap}). Since Galactic foregrounds are strongly polarized, whereas the tSZ signal is largely unpolarized, we propose and explore the incorporation of Stokes  $Q$ and $U$ polarization maps alongside total intensity maps within the NILC framework to enhance Galactic foreground mitigation in the reconstructed tSZ $y$-map. This multi-Stokes extension of NILC, first outlined in \cite{Remazeilles2025}, leverages the intrinsic $TE$ and $TB$ correlations of thermal dust and synchrotron emissions, providing an additional discriminant—polarization—beyond frequency information alone to separate Galactic foregrounds from the extragalactic tSZ signal.

At small angular scales, $y$-maps are primarily affected by residual CIB contamination. Unlike Galactic foregrounds, the CIB is largely unpolarized, and therefore the inclusion of polarization channels may only have a limited impact on its mitigation. Although external LSS tracers correlated with the CIB could, in principle, be incorporated into NILC as auxiliary channels to improve CIB removal, as proposed in \citep{Kusiak2023} for CMB reconstruction, such an approach is less suitable for tSZ reconstruction, since the signal of interest is itself correlated with these tracers. As the cumulative emission of dusty star-forming galaxies over cosmic history, the CIB not only exhibits a non-trivial effective SED due to the superposition of redshifted modified blackbody spectra \citep{Knox2001, Bethermin2013, Bethermin2015}, but is also spatially correlated with the tSZ signal \citep{Addison2012, George2015, planck2014-a29, Maniyar2021}, as galaxy clusters hosting hot gas also host the galaxies responsible for the CIB. This correlation between signal (tSZ) and foregrounds (CIB) thus requires explicit CIB deprojection using Constrained ILC (CILC) techniques \citep{2011MNRAS.410.2481R, Remazeilles2021, NILC-McCarthyHill, PhysRevD.109.063530}, rather than simple variance minimization with NILC, in order to avoid reconstruction biases and spurious correlations in cross-correlation studies of tSZ with other LSS tracers, such as CMB lensing \citep{Hill2014, 2015A&A...575L..11H, McCarthy2024b} and galaxy surveys \citep{2017MNRAS.467.2315V, Makiya2018, Pandey2019, Koukoufilippas2020, Yan2021}.

The complex SED of the CIB can be modelled through a moment expansion \citep{Chluba2017} and mitigated in tSZ reconstruction via moment deprojection within the CILC framework \citep{Remazeilles2021}. Recent efforts have applied CIB moment deprojection to tSZ cluster detection using constrained multi-matched filters \citep{2023MNRAS.522.5123Z} and to $y$-map reconstruction using CILC methods \citep{NILC-McCarthyHill}. However, imposing additional deprojection constraints, while reducing CIB-induced biases, generally increases the noise variance of the reconstructed tSZ signal, particularly when the number of available frequency channels is limited. To balance CIB mitigation against noise amplification, in this work, we explore a selective CIB moment deprojection strategy in which only specific CIB moments are deprojected, while the remaining components are treated through blind variance minimization.

This work is organised as follows. Section~\ref{sec:method} introduces our new foreground-removal strategies for thermal SZ $y$-map reconstruction, including a multi-Stokes Hybrid NILC approach that combines temperature and polarization channels to mitigate Galactic foreground contamination (Section~\ref{sec:TP}), and a selective CIB moment deprojection scheme using CILC to optimise CIB removal against noise penalty (Section~\ref{sec:deproj}). Section~\ref{sec:data} describes the \pl\ PR4 data and simulations used throughout the analysis. Results and validation based on both simulations and data are presented in Section~\ref{sec:results}. We discuss the implications for future surveys and present our conclusions in Section~\ref{sec:conclusions}. Additional validation tests and comparisons with publicly available $y$-maps are provided in Appendices~\ref{app:QU_choice} and \ref{app:cib_deproj_comparison}.

%%%%%%%%%%%%%%%%%%%%%%%%%%%%%%%%%%%%%%%%%%%%%%%%%%%%%%%%%%%%%%%%%%%%%%%%%%%%%%%%%%%%%%%%%%%%%%%%%%%%
\section{Foreground removal strategies for SZ maps} \label{sec:method}

The thermal SZ effect has a distinct and well-understood SED in the non-relativistic limit \citep{1969SunyaevZeldovich, 1972SunyaevZeldovich},
\begin{align}
\label{eq:sed}
    g_{\nu} = T_{\rm CMB} \left[ x \coth\left(\frac{x}{2}\right) - 4 \right] \,, \quad \text{with} \quad x = \frac{h\nu}{k_B T_{\rm CMB}}\,,
\end{align}
where $h$ is the Planck constant, $k_B$ is the Boltzmann constant, and $T_{\rm CMB} = 2.7255$\,K is the CMB temperature. This distinctive SED enables the separation of the thermal SZ signal from other astrophysical components in multi-frequency observations.

The multi-frequency data $x_\nu(p)$, in either temperature or polarization, observed at frequencies $\nu$ and sky pixels $p$, can be written as
\begin{align}
	\label{eq:model}
    	x_{\nu}( p) = \varepsilon g_{\nu}y(p) + n_{\nu}(p)\,, 
\end{align}
where $g_{\nu}y(p)$ denotes the thermal SZ contribution, with a spatially varying Compton parameter $y(p)$ and a spatially uniform SED $g_{\nu}$ across the sky. The term $n_{\nu}(p)$ represents the \emph{unparametrized} contamination from instrumental noise, CMB, and astrophysical foregrounds. The parameter $\varepsilon$ equals unity for temperature (total-intensity) data and vanishes for polarization data, reflecting the fact that the tSZ effect is unpolarized.

In contrast to the tSZ signal, the SED of Galactic foregrounds, mainly thermal dust and synchrotron emission, and that of extragalactic foregrounds, mainly the CIB, vary spatially and are not fully characterised. This motivates the use of \emph{blind} component separation methods, such as the internal linear combination (ILC), which are well suited for extracting a minimum-variance estimate of the thermal SZ Compton-$y$ parameter, $y(p)$, from the multifrequency data $x_{\nu}( p)$ without relying on explicit foreground modelling.

\subsection{Multi-Stokes hybrid ILC combining temperature and polarization for enhanced Galactic foreground mitigation}
\label{sec:TP}

The multi-Stokes hybrid ILC method (hereafter, Hybrid ILC) was first introduced in \cite{Remazeilles2025}, where $E$- and $B$-mode frequency maps were jointly combined and optimally weighted within NILC to disentangle the component of the CMB $E$-mode anisotropies uncorrelated with $B$-modes from that correlated with $B$-modes through cosmic birefringence. In this framework, the signal channels (the $E$-mode frequency maps) are weighted to preserve the targeted component (the uncorrelated CMB $E$-mode), while the auxiliary channels (the $B$-mode frequency maps), which contain no contribution from the target, are weighted to suppress the contaminant that appears correlated across both signal and auxiliary channels (the part of CMB $E$-modes correlated with $B$-modes through cosmic birefringence).

The Hybrid ILC framework is not limited to cosmic birefringence and can be applied in other contexts. In this work, we adopt it to enhance the mitigation of Galactic foreground contamination in unpolarized extragalactic signals, specifically the thermal SZ effect. In this case, the signal channels correspond to the temperature (or total-intensity) frequency maps, while the auxiliary channels are the Stokes $Q$ and $U$ polarization maps. This framework exploits the intrinsic $TE$ and $TB$ correlations of synchrotron and thermal dust emission, providing additional leverage to reduce residual Galactic foreground contamination in the reconstructed SZ map.

To simplify the presentation without loss of generality, we consider here the combination of temperature (total-intensity) frequency maps and $E$-mode polarization frequency maps. The formalism can be straightforwardly extended to include $B$-mode maps or to work with Stokes $Q$ and $U$ fields instead of $E$ and $B$.

\subsubsection{Derivation of the Hybrid ILC weights}
\label{sec:TPweights}

The Hybrid ILC estimate of the thermal SZ Compton-$y$ field, using both $n_T$ temperature ($T$) channels and $n_E$ polarization ($E$) channels\footnote{We consider the $E$ field here, but the derivation applies equally to other polarization fields such as $B$, $Q$, $U$, or $P$.}, is constructed as the weighted linear combination
\begin{align}
\hat{y}(p) &= \bdw^t \bdx(p) \cr
        & =  \begin{pmatrix}
         \bdw_{T}^t\ \bdw_{E}^t \end{pmatrix}  \begin{pmatrix}
         \bdt(p)\\ \bde(p) \end{pmatrix} \cr 
         & = \bdw_{T}^t \bdt(p) +  \bdw_{E}^t \bde(p) \cr
         & = \sum_{\nu=1}^{n_T}w_{T,\nu}\,  t_{\nu}(p) + \sum_{\nu=1}^{n_E}w_{E,\nu}\,  e_{\nu}(p)\,,
\end{align}
where $t_\nu(p)$ and $e_\nu(p)$ denote the temperature and polarization (here, $E$-mode) frequency maps, respectively, $w_{T,\nu}$ and $w_{E,\nu}$ are the corresponding ILC weights, and bold symbols denote vectors. 

The variance of the Hybrid ILC estimate,
\begin{align}
\label{eq:variance}
\langle \hat{y}^2(p) \rangle &= \bdw^t \langle \bdx(p)  \bdx^t (p)\rangle \bdw \cr &= \bdw^t  \tC(p) \bdw \cr
&= \begin{pmatrix} \bdw_{T}^t\ \bdw_{E}^t \end{pmatrix} 
 \begin{pmatrix}  
         \tC^{TT}\  \tC^{TE}\\  \tC^{ET}\  \tC^{EE}
         \end{pmatrix}  (p)
         \begin{pmatrix}  \bdw_{T}\\ \bdw_{E} \end{pmatrix} \cr
         &=  \bdw_{T}^t  \tC^{TT}(p) \bdw_{T} + \bdw_{E}^t  \tC^{EE}(p) \bdw_{E} + \bdw_{T}^t  \tC^{TE}(p) \bdw_{E} + \bdw_{E}^t  \tC^{ET}(p) \bdw_{T}  \,,
\end{align}
is minimized subject to the constraint
\begin{align}
\label{eq:constraint}
\bdw^t \bda_{\rm tSZ} =  \begin{pmatrix}
         \bdw_{T}^t\ \bdw_{E}^t \end{pmatrix}  \begin{pmatrix}
         \bdg\\ \boldsymbol{0} \end{pmatrix} =  \bdw_{T}^t  \bdg = \sum_{\nu=1}^{n_T} w_{T,\nu}\,   g_{\nu} = 1\,,
\end{align}
ensuring that the full thermal SZ signal is preserved during variance minimization. Here, 
\begin{align}
\label{eq:effsed}
\bda_{\rm tSZ} =  \begin{pmatrix} \bdg\\ \boldsymbol{0} \end{pmatrix}
\end{align} 
is the effective SED vector of the tSZ signal across all channels, with the subvector $\bdg$ collecting the tSZ SED coefficients $g_{\nu}$ across the temperature channels, while the SED coefficients for the polarization channels are set to zero since the tSZ signal is unpolarized. The ${(n_T+n_E) \times (n_T+n_E)}$ matrix $\tC(p)$ describes the full data covariance matrix across all pairs of channels, while its blocks $\tC^{TT}(p)$  and $\tC^{EE}(p)$  correspond to the temperature and polarization covariance matrices, respectively, and $\mathbf{C}^{TE}(p)$ captures the cross-covariance between temperature and polarization channels.\footnote{Note that $\mathbf{C}^{ET}=\left(\mathbf{C}^{TE}\right)^t\neq \mathbf{C}^{TE}$, since the covariance between $t_{\nu_1}(p)$ and $e_{\nu_2}(p)$ differs from that between $t_{\nu_2}(p)$ and $e_{\nu_1}(p)$ for a pair of distinct frequencies $(\nu_1,\nu_2)$.}

The Hybrid ILC weights $\bdw$ can be obtained by minimizing the Lagrangian
\begin{align}
\label{eq:lagrange}
\mathcal{L} &= \bdw^t  \tC \bdw + \lambda\left(1 - \bdw^t \bda_{\rm tSZ}\right)\cr
\cr
&=  \bdw_{T}^t  \tC^{TT} \bdw_{T} + \bdw_{E}^t  \tC^{EE} \bdw_{E} + \bdw_{T}^t  \tC^{TE} \bdw_{E} + \bdw_{E}^t  \tC^{ET} \bdw_{T} +\, \lambda\left(1 - \bdw_{T}^t  \bdg \right)\,,
\end{align}
where the Lagrange multiplier $\lambda$ enforces the constraint Eq.~\eqref{eq:constraint}. Minimizing the first line of Eq.~\eqref{eq:lagrange} yields
\begin{align}
\label{eq:hybridweights}
\bdw = \frac{ \tC^{-1}\bda_{\rm tSZ}}{ \bda_{\rm tSZ}^t \tC^{-1}\bda_{\rm tSZ} }\,, 
\end{align}
which resembles the standard expression for the ILC weights, but now implicitly incorporates both temperature and polarization information. Equation~\eqref{eq:hybridweights} is the form of the Hybrid ILC weights used in practice. However, it is instructive to understand how these weights operate differently on the temperature versus polarization channels, as we show hereafter.

To derive explicit temperature and polarization weights, $\bdw_{T}$ and $\bdw_{E}$, one can either minimize the second-line expression of the Lagrangian in Eq.~\eqref{eq:lagrange}, as in \cite{Remazeilles2025}, or equivalently compute the inverse covariance matrix $\tC^{-1}$ in Eq.~\eqref{eq:hybridweights} via $2\times 2$ block matrix inversion.  Defining the matrix
\begin{align}
\label{eq:pearson}
\tP = \left(\tC^{TT}\right)^{-1}\tC^{TE} \left(\tC^{EE}\right)^{-1}\tC^{ET}\,,
\end{align}
which quantifies the effective degree of correlation between temperature and polarization channels, the inverse covariance matrix reads
\begin{align}
\label{eq:inverse_cov}
\tC^{-1} = \begin{pmatrix}  \tS^{-1} & \qquad -  \tS^{-1} \tC^{TE} \left(  \tC^{EE} \right)^{-1} \\[0.2cm]  - \left(  \tC^{EE} \right)^{-1}\tC^{ET}   \tS^{-1}  & \qquad  \left(  \tC^{EE} \right)^{-1}  + \left(  \tC^{EE} \right)^{-1}\tC^{ET} \tS^{-1}\tC^{TE}\left(  \tC^{EE} \right)^{-1}  \end{pmatrix}\,,
\end{align}
where
\begin{align}
\label{eq:schur}
\tS = \tC^{TT}\left(\tI - \tP\right)\,
\end{align}
is the Schur complement. Substituting Eqs.~\eqref{eq:inverse_cov} and \eqref{eq:effsed} into Eq.~\eqref{eq:hybridweights} yields 
\begin{align}
\label{eq:hybridweights_te}
\bdw = \begin{pmatrix}  \bdw_{T} \\[0.2cm] \bdw_{E} \end{pmatrix} = \begin{pmatrix} \frac{  \tS^{-1}\bdg}{  \bdg^t\tS^{-1}\bdg} \\[0.2cm] - \frac{ \left(\tC^{EE}\right)^{-1}\tC^{ET} \tS^{-1}\bdg}{  \bdg^t\tS^{-1}\bdg} \end{pmatrix}\,.
\end{align}
Using Eq.~\eqref{eq:schur}, the temperature and polarization weights can finally be rewritten as:
\begin{align}
\label{eq:weights_t}
 \bdw_{T}  &= \frac{  \left(\tI - \tP\right)^{-1}\left(\tC^{TT}\right)^{-1}\bdg }{  \bdg^t \left(\tI - \tP\right)^{-1}\left(\tC^{TT}\right)^{-1}\bdg }\,, \\
 \label{eq:weights_e}
\bdw_{E} &=   - \left(\tC^{EE}\right)^{-1}\tC^{ET}\, \bdw_{T} \,.
\end{align}

If there were no correlation between the temperature and polarization channels, i.e. $\tC^{TE}=\tC^{ET}=0$ and thus ${\tP=0}$, then the Hybrid ILC weights (Eqs.~\ref{eq:weights_t}-\ref{eq:weights_e}) would reduce to the standard ILC operating on temperature channels, with $\bdw_{T}  = ( \bdg^t (\tC^{TT})^{-1}\bdg )^{-1} (\tC^{TT})^{-1}\bdg $ and $\bdw_{E} = 0$.  However, such correlations exist for Galactic foregrounds, though not for the tSZ signal. As a result, the Hybrid ILC weights in Eqs.~\eqref{eq:weights_t}-\eqref{eq:weights_e} allow for additional suppression of Galactic foreground variance compared to the standard ILC, while still preserving the targeted tSZ signal, as we demonstrate in the next section.

\subsubsection{Variance suppression factor}
\label{sec:TPvar}

Using the expression for the polarization weights $\bdw_{E}$ (Eq.~\ref{eq:weights_e}), the variance of the Hybrid ILC estimate, $\hat{y}$, of the tSZ Compton parameter (Eq.~\ref{eq:variance}) can be rewritten in terms of the temperature weights $\bdw_T$ as
\begin{align}
\label{eq:hybridvar0}
\langle \hat{y}^2 \rangle &= \bdw_{T}^t  \tC^{TT} \bdw_{T} + \bdw_{E}^t  \tC^{EE} \bdw_{E} + \bdw_{T}^t  \tC^{TE} \bdw_{E}  + \bdw_{E}^t  \tC^{ET} \bdw_{T}\cr
  &=  \bdw_{T}^t  \tC^{TT} \bdw_{T} -  \bdw_{T}^t  \tC^{TE} \left(\tC^{EE}\right)^{-1}\tC^{TE}\bdw_{T}  \cr
  &=  \bdw_{T}^t  \tC^{TT}\left(\tI - \tP\right)\bdw_{T}\,,
\end{align}
where $\tP$ is the correlation matrix defined in Eq.~\eqref{eq:pearson}. 

Substituting the expression for the temperature weights $\bdw_{T}$ (Eq.~\ref{eq:weights_t}) then gives
\begin{align}
\label{eq:hybridvar}
\langle \hat{y}^2 \rangle &=  \frac{1}{\bdg^t \left(\tI - \tP\right)^{-1}\left(\tC^{TT}\right)^{-1}\bdg}\,,
\end{align}
which contrasts with the variance resulting from a standard ILC operating on temperature channels only:
\begin{align}
\label{eq:stdvar}
\langle \hat{y}^2_{\rm std} \rangle = \bdw_{T,{\rm std}}^t  \tC^{TT} \bdw_{T,{\rm std}} =  \frac{1}{\bdg^t \left(\tC^{TT}\right)^{-1}\bdg}\,,
\end{align}
where the standard ILC weights, $\bdw_{T,{\rm std}} = ( \bdg^t (\tC^{TT})^{-1}\bdg )^{-1} (\tC^{TT})^{-1}\bdg $, correspond to the special case of the Hybrid ILC weights with $\tP = 0$.

Since the matrix $\tP$ (Eq.~\ref{eq:pearson}) measures the squared, normalized correlation between temperature and polarization channels, its eigenvalues lie in the interval $[0,1]$, and therefore the eigenvalues of the inverse matrix $(\tI - \tP)^{-1}$ are in turn larger than $1$. As a result,  for any vector $\bdg$, we have $\bdg^t (\tI - \tP)^{-1}(\tC^{TT})^{-1}\bdg > \bdg^t (\tC^{TT})^{-1}\bdg$, so that the Hybrid ILC variance (Eq.~\ref{eq:hybridvar}) is smaller than the standard ILC variance (Eq.~\ref{eq:stdvar}). Therefore, including polarization channels that are correlated with the temperature channels in a Hybrid ILC reduces the overall variance of the reconstructed $y$-map by a factor
\begin{align}
\label{eq:varratio}
 \frac{ \langle \hat{y}^2 \rangle }{ \langle \hat{y}^2_{\rm std} \rangle } =  \frac{\bdg^t \left(\tC^{TT}\right)^{-1}\bdg}{\bdg^t \left(\tI - \tP\right)^{-1}\left(\tC^{TT}\right)^{-1}\bdg} < 1\,
\end{align}
relative to the variance of the $y$-map obtained from a standard ILC using temperature channels only.

This variance reduction in the reconstructed $y$-map (Eq.~\ref{eq:varratio}) applies exclusively to Galactic foreground (and CMB) contamination, which is correlated between temperature and polarization, while the tSZ variance is fully preserved. This can be shown by decomposing the covariance matrices into tSZ and non-tSZ contributions, leading to
\begin{align}
\label{eq:split1}
\tC^{TT} = \langle y^2 \rangle\bdg\bdg^t + \tN^{TT}  \,,
\end{align}
where $\tN^{TT}$ is the covariance matrix of the foregrounds and noise in temperature, while $ \langle y^2 \rangle\bdg\bdg^t$ is the covariance matrix of the tSZ signal. In contrast, we have $\tC^{EE} = \tN^{EE}$ and $\tC^{TE} = \tN^{TE}$ since there is no tSZ contribution in polarization channels. Therefore,
\begin{align}
\label{eq:split2}
\tC^{TT}\left(\tI - \tP\right) = \langle y^2 \rangle\bdg\bdg^t + \tN^{TT}\left(\tI - \tP_{\tN}\right)\,,
\end{align}
where the matrix
\begin{align}
\label{eq:fgpearson}
\tP_{\tN} = \left(\tN^{TT}\right)^{-1}\tN^{TE} \left(\tN^{EE}\right)^{-1}\tN^{ET}\,
\end{align}
 quantifies the degree of correlation between temperature and polarization due to non-tSZ components, in particular Galactic foregrounds. 

Applying the Woodbury formula, $(A+UBV)^{-1}=A^{-1}-A^{-1}U(B^{-1}+VA^{-1}U)^{-1}VA^{-1}$, with $A=\tN^{TT}$, $B=\langle y^2 \rangle$, $U=\bdg$, and $V=\bdg^t$, gives
\begin{align}
\left(\tC^{TT}\right)^{-1} = \left(\langle y^2 \rangle\bdg\bdg^t + \tN^{TT} \right)^{-1} = \left(\tN^{TT}\right)^{-1} - \frac{\left(\tN^{TT}\right)^{-1}\bdg \bdg^t \left(\tN^{TT}\right)^{-1}}{\frac{1}{\langle y^2 \rangle} +  \bdg^t \left(\tN^{TT}\right)^{-1}\bdg}\,,
\end{align}
so that the variance of the standard ILC $y$-map (Eq.~\ref{eq:stdvar}) becomes
\begin{align}
\label{eq:stdvarbis}
\langle \hat{y}^2_{\rm std} \rangle = \langle y^2 \rangle + \frac{1}{\bdg^t \left(\tN^{TT}\right)^{-1}\bdg}\,.
\end{align}

Similarly, applying the Woodbury formula to invert the matrix in Eq.~\eqref{eq:split2}, with $A = \tN^{TT}(\tI - \tP_{\tN})$ instead of $A = \tN^{TT}$, yields the variance of the Hybrid ILC $y$-map (Eq.~\ref{eq:hybridvar}) as
\begin{align}
\label{eq:hybridvarbis}
\langle \hat{y}^2 \rangle = \langle y^2 \rangle + \frac{1}{\bdg^t \left(\tI - \tP_{\tN}\right)^{-1}\left(\tN^{TT}\right)^{-1}\bdg}\,.
\end{align}
Hence, the variance of the tSZ signal itself, $\langle y^2 \rangle$, is fully preserved by the Hybrid ILC in Eq.~\eqref{eq:hybridvarbis}, while the variance of the foregrounds is further reduced relative to the standard ILC by the factor
\begin{align}
\label{eq:varratiobis}
  \frac{\bdg^t \left(\tN^{TT}\right)^{-1}\bdg}{\bdg^t \left(\tI - \tP_{\tN}\right)^{-1}\left(\tN^{TT}\right)^{-1}\bdg} < 1\,.
\end{align}

As evident from Eq.~\eqref{eq:varratiobis}, the stronger the correlation between temperature and polarization channels in $\tP_{\tN}$, the greater the reduction of Galactic foreground variance in the Hybrid ILC $y$-map. However, $\tN^{TT}$ and $\tN^{EE}$ are not purely foreground covariance matrices, but also include instrumental noise, which may attenuate the measured temperature-polarization correlation in $\tP_{\tN}$ (Eq.~\ref{eq:fgpearson}) relative to the intrinsic correlation of the Galactic foregrounds (see the right panel of Figure~\ref{fig:eigvals}, shown later, which compares the eigenvalues of the temperature-polarization correlation matrix $\tP$ for different noise levels). Consequently, lower instrumental noise enhances the effectiveness of the multi-Stokes Hybrid ILC, allowing future high-sensitivity experiments such as \lb\ or SO to exploit its full potential far better than \pl.

Finally, the formalism presented here is independent of the domain in which the Hybrid ILC is implemented. While the derivation above assumes the pixel domain (as indicated by the pixel index $p$ in the equations), it applies equally to the spherical harmonic or needlet domains.

\subsection{Selective moment deprojection with CILC for optimized CIB subtraction}
\label{sec:deproj}

\subsubsection{CIB moment expansion}
\label{sec:cibmoment}

As the cumulative emission from dusty star-forming galaxies throughout cosmic history, including those hosted by galaxy clusters detected via thermal SZ effect, the CIB constitutes a significant extragalactic foreground for thermal SZ $y$-maps, particularly at small angular scales. Since the CIB is expected to have negligible polarized emission \citep{2020ApJ...897..140F}, the inclusion of polarization channels, as described in Section~\ref{sec:TP}, has only a minor impact on residual CIB contamination in the reconstructed $y$-map. Although improving Galactic foreground removal with the Hybrid NILC can indirectly grant the ILC additional freedom to minimize CIB variance in the temperature channels, this effect is expected to remain secondary. Consequently, an additional foreground-removal step based on spectral deprojection using CILC \citep{2011MNRAS.410.2481R, Remazeilles2021} must be applied to efficiently mitigate CIB, in addition to the multi-Stokes Hybrid ILC used to suppress Galactic foregrounds.

The CIB contribution from dusty galaxies at a redshift $\bar{z}$ can be modelled as a modified blackbody (MBB) emission with a redshifted temperature,
\begin{align}
\label{eq:mbb}
I_\nu(\bar{z},\bar{\beta},\bar{T}) \propto \left[ \nu(1+\bar{z}) \right]^{\bar{\beta}}
B_{\nu(1+\bar{z})}\!\left(\bar{T}(\bar{z}) = T_0(1+\bar{z})^\alpha \right)\,,
\end{align}
where $\bar{\beta} \simeq 1.75$, $T_0 \simeq 24.4\,\mathrm{K}$, and $\alpha \simeq 0.36$ \citep{planck2013-pip56}, and $B_\nu(T)$ denotes the Planck's blackbody function.
The Planck function $B_{\nu(1+\bar{z})}(T)$ in Equation~\eqref{eq:mbb} is evaluated at the source rest-frame frequency $\nu(1+\bar{z})$. Since the Planck spectrum depends on the combination $h\nu/kT$, this convention is exactly equivalent to expressing the blackbody in terms of the observed frequency $\nu$ with an effective temperature $T/(1+\bar{z})$. Hence, no additional temperature rescaling is required in our formulation.

As the observed CIB signal is the cumulative emission integrated over redshift, the total CIB intensity at frequency $\nu$ and pixel $p$ is the average of multiple MBB emitters over redshifts:
\begin{align}
\label{eq:meanz}
I^{\rm CIB}_\nu(p) = \int I_\nu(z,\beta,T)\, dz \propto \left\langle I_\nu(z,\beta,T) \right\rangle_z\,,
\end{align}
which implies that the effective CIB SED deviates from a simple MBB form. These deviations can be described through a Taylor expansion of the MBB SED in terms of moments of the temperature and spectral index around pivot values \citep{Chluba2017}:
\begin{align}
\label{eq:cib}
I^{\rm CIB}_\nu(p) &\propto \left\langle I_\nu(z,\beta,T) \right\rangle_z \cr
& = I_\nu(\bar{z},\bar{\beta},\bar{T}) +  \left\langle\beta - \bar{\beta}\right\rangle_z\frac{\partial I_\nu}{\partial\bar{\beta}}(\bar{z},\bar{\beta},\bar{T})+   \left\langle T - \bar{T}\right\rangle_z\frac{\partial I_\nu}{\partial\bar{T}}(\bar{z},\bar{\beta},\bar{T})   \cr
& + \frac{1}{2} \left\langle\left(\beta - \bar{\beta}\right)^2\right\rangle_z\frac{\partial^2 I_\nu}{\partial\bar{\beta}^2}(\bar{z},\bar{\beta},\bar{T}) +  \frac{1}{2} \left\langle\left( T - \bar{T}\right)^2\right\rangle_z\frac{\partial^2 I_\nu}{\partial\bar{T}^2}(\bar{z},\bar{\beta},\bar{T}) \cr
&+  \left\langle\left(\beta - \bar{\beta}\right)\left( T - \bar{T}\right)\right\rangle_z\frac{\partial^2 I_\nu}{\partial\bar{\beta}\partial\bar{T}}(\bar{z},\bar{\beta},\bar{T}) +\mathcal{O}(\beta^3,T^3)\,.
\end{align}

Since the thermal SZ signal targeted in this work originates primarily from \pl\ galaxy clusters with a mean redshift $\bar{z} \simeq 0.2$, we adopt this value to define the CIB pivot parameters, setting $\bar{T} = 26.1\,\mathrm{K}$ and $\bar{\beta} = 1.75$ for the moment expansion in Equation~\eqref{eq:cib}. This choice is motivated by our intent to suppress the CIB component that is most strongly correlated with the tSZ signal, rather than by the redshift at which the total CIB emissivity peaks, as this CIB emission is sourced by galaxies hosted by, or associated with, the same halos that generate the tSZ signal. For applications such as stacking the reconstructed $y$-map on clusters, this correlated CIB component is most likely to bias the recovered tSZ signal if not properly suppressed. The expansion in Equation~\eqref{eq:cib} accounts not only for spectral parameter variations across redshift but also for spatial variations across the sky, with $\beta = \beta(p)$ and $T = T(z,p)$.

\subsubsection{CIB moment deprojection}
\label{sec:cibcilc}

Each CIB moment, e.g.\ $\langle (T - \bar{T})^k \rangle_z / k!$, can be interpreted as an effective foreground component with a spatially uniform SED given by $\partial^k I_\nu(\bar{z},\bar{\beta},\bar{T}) / \partial \bar{T}^k$. Each of these components can, in principle, be deprojected using CILC by imposing nulling constraints on the corresponding SEDs, i.e.
\begin{equation}
\sum_\nu w_\nu\, \frac{\partial^k I_\nu}{\partial \bar{T}^k} = 0\,,
\end{equation}
where $w_\nu$ are the CILC weights across frequency channels, and thus form a vector $\bdw$ that is orthogonal to the vector collecting the moment SED coefficients.
The resulting CILC weights are given by \citep{Remazeilles2021}
\begin{align}
\label{eq:wcilc}
\bdw^t = \bde^t \left( \tA^t \tC^{-1} \tA \right)^{-1} \tA^t \tC^{-1}\,,
\end{align}
where the SED mixing matrix $\tA$ is, in this example, an $n_T \times 2$ matrix,\footnote{In the more general multi-Stokes Hybrid ILC framework introduced in Section~\ref{sec:TP}, the mixing matrix generalizes, in this example, to a $(n_T + n_E) \times 2$ matrix with null entries in the polarization block,
\begin{align}
\label{eq:sedmatrix2}
\tA =
\begin{pmatrix}
\bdg & \,\partial^k \boldsymbol{I} / \partial \bar{T}^k\\
\boldsymbol{0} & \,\boldsymbol{0}
\end{pmatrix}\,.
\end{align}}
\begin{align}
\label{eq:sedmatrix}
\tA =
\begin{pmatrix}
\bdg & \,\partial^k \boldsymbol{I} / \partial \bar{T}^k
\end{pmatrix},
\end{align}
with the column vector $\bdg$ collecting the thermal SZ SED coefficients $g_\nu$, and the column vector $\partial^k \boldsymbol{I} / \partial \bar{T}^k$ the SED coefficients $\partial^k I_\nu(\bar{z},\bar{\beta},\bar{T})/\partial\bar{T}^k$ of the $k^{\rm th}$-order CIB moment $\langle (T - \bar{T})^k \rangle_z / k!$. The vector $\bde$ selects the thermal SZ component and is defined as
\begin{align}
\bde^t =
\begin{pmatrix}
1 & 0
\end{pmatrix}.
\end{align}

This approach, known as \emph{moment deprojection} (or cMILC, for constrained moment ILC), was introduced in \citep{Remazeilles2021} and first applied to CMB $B$-mode polarization. It has since been applied to thermal SZ $y$-map reconstruction \citep{NILC-McCarthyHill, PhysRevD.109.063530}, although with selected moments and pivot parameters that differ from those adopted in this work.

In practice, increasing the number of deprojected CIB moments in CILC leads to a higher variance contribution from instrumental noise and unconstrained foregrounds, including unconstrained higher-order CIB moments. This is due to the finite number of available frequency channels, as each additional nulling constraint effectively reduces the degrees of freedom available for variance minimization. As emphasized by \citep{Carones2024}, this introduces a bias--variance trade-off, motivating a strategy of \emph{selective} moment deprojection rather than blindly deprojecting all moments in the Taylor expansion, since some of these moments might be better suited for variance minimization when they significantly contribute to the overall variance. 

Following this approach, we explore several CILC configurations involving selective deprojection of zeroth- and first-order CIB moments, summarized in Table~\ref{tab:cilc}. The resulting $y$-maps will be compared in Section~\ref{sec:results} in terms of residual CIB contamination, quantified through cross-power spectra with the \pl\ GNILC CIB map \citep{planck2016-XLVIII}, and in terms of the overall variance penalty, assessed via the $y$-map auto-power spectrum. 
In this analysis, second-order CIB moments are not deprojected but instead left to blind variance minimization, as their associated noise penalty is too large given the sensitivity and frequency coverage of \pl.

\begin{table}[t]
\centering
    \resizebox{\columnwidth}{!}{
\renewcommand{\arraystretch}{1.3}
\begin{tabular}{lll}
\hline
\textbf{Method name}  &  \textbf{Deprojected CIB moments} & \textbf{Associated SEDs} \\
\hline
CILC CIB$[0]$   &  zeroth-order  &  $I_\nu$ \\
CILC CIB$[0,d\beta]$  &  zeroth-order, first-order in $\beta$  &  $I_\nu$, $\frac{\partial I_\nu}{\partial\bar{\beta}}$ \\
CILC CIB$[0,d\beta,dT]$   &  zeroth-order, first-order in $\beta$ and $T$  &  $I_\nu$, $\frac{\partial I_\nu}{\partial\bar{\beta}}$, $\frac{\partial I_\nu}{\partial\bar{T}}$ \\
CILC CIB$[0,dT]$  &  zeroth-order, first-order in $T$  &  $I_\nu$, $\frac{\partial I_\nu}{\partial\bar{T}}$ \\
CILC CIB$[0,dT]\, (j\geq 6)$  &  zeroth-order for $j\geq 6$ , first-order in $T$ for $j\geq 6$  &  $I_\nu$ for $j\geq 6$, $\frac{\partial I_\nu}{\partial\bar{T}}$ for $j\geq 6$ \\
CILC CIB$[0,d\beta(j < 6),dT(j\geq 6)]$  &  zeroth-order, first-order in $\beta$ for $j <6$ and $T$ for $j\geq 6$  &  $I_\nu$, $\frac{\partial I_\nu}{\partial\bar{\beta}}$ for $j < 6$, $\frac{\partial I_\nu}{\partial\bar{T}}$ for $j\geq 6$  \\
\hline
\end{tabular}}
\caption{CILC configurations for CIB moment deprojection in the $y$-map. The notation $j < 6$ (resp.\ $j \geq 6$) indicates that the deprojection constraint is applied only to the first five needlet bands 1--5 (lower multipoles) or to the last five needlet bands 6--10 (higher multipoles), respectively.}
\label{tab:cilc}
\end{table}

As we will show in Section~\ref{sec:results}, leaving the first-order CIB moment in $\beta$ to blind variance minimization while deprojecting only the first-order temperature moment (i.e., CILC CIB$[0,dT]$ in Table~\ref{tab:cilc}) can provide a more optimal balance between residual CIB contamination and noise amplification.

%%%%%%%%%%%%%%%%%%%%%%%%%%%%%%%%%%%%%%%%%%%%%%%%%%%%%%%%%%%%%%%%%%%%%%%%%%%%%%%%%%%%%%%%%%%%%%%%%%%%
\section{Data and simulations}
\label{sec:data}

\subsection{\pl\ Release 4 (PR4) data}
\label{sec:PR4}

We use the \pl\ Release 4 (PR4) data \citep{planck2020-LVII} to reconstruct thermal SZ Compton-$y$ maps with our component separation pipeline and to assess the performance of the new foreground-removal strategies introduced in this work on real observations. The PR4 data consist of nine calibrated full-sky frequency maps in total intensity (Low Frequency Instrument, LFI: $30$--$70$\,GHz; High Frequency Instrument, HFI: $100$--$857$\,GHz) and seven frequency maps in Stokes $Q$ and $U$ polarization (LFI: $30$--$70$\,GHz; HFI: $100$--$353$\,GHz). Compared to previous \pl\ data releases, PR4 includes several important improvements, such as reduced noise through $\sim 8$\,\% additional repointing data; refined 4\,K line fitting; improved pixel flagging and glitch removal, leading to lower half-ring correlations; destriping with \textsc{Madam} using very short baselines to suppress scan-direction stripes; and a coherent LFI and HFI calibration with updated bandpasses.

In addition to the full-mission maps, PR4 half-ring data splits (HR1 and HR2) are used to characterize the noise properties of the reconstructed full-mission $y$-maps. They are also employed to obtain a noise-debiased thermal SZ power spectrum via the cross-power spectrum of the reconstructed HR1 and HR2 $y$-maps. Details of preprocessing and beam treatment are provided in Section~\ref{sec:NILC_implementation}.

The PR4 polarization data are exploited in this work to further reduce Galactic foreground contamination in the reconstructed thermal SZ $y$-maps by combining temperature and polarization frequency maps within the NILC framework (see Section~\ref{sec:TP}).

\subsection{Simulations}
\label{sec:sims}

As a proof-of-concept study, we also apply our methods to two sets of sky simulations: \pl-\emph{like} simulations and \emph{low-noise} simulations with a sensitivity improved by a factor of $25$ relative to \pl, comparable to that expected for future CMB experiments such as \lb. This allows us to demonstrate how the effectiveness of the multi-Stokes Hybrid NILC approach developed in this work improves with increasing instrumental sensitivity. In particular, higher sensitivity reduces the noise-induced attenuation of the temperature--polarization correlation of Galactic foregrounds, thereby revealing the full potential of the Hybrid NILC method for upcoming CMB surveys.

We use the \emph{Planck Sky Model} (PSM; \cite{PSM2013}) to simulate microwave sky observations at \pl\ frequencies in both total intensity and polarization, adopting either nominal \pl\ channel sensitivities or enhanced sensitivities improved by a factor of $25$. In total intensity, the simulated extragalactic components include the CMB, thermal and kinetic SZ effects, the CIB, and radio and infrared compact sources, while the Galactic components include thermal dust, synchrotron, free-free emission, and anomalous microwave emission. In polarization, the simulated extragalactic components include only the CMB and compact sources, while the Galactic components are limited to thermal dust and synchrotron.

The templates and SED models adopted for all components are identical to those used in Section~3.2 of \cite{Remazeilles2022}, to which we refer the reader for a comprehensive description. Below, we briefly summarize the components most relevant to this study.

The thermal SZ signal is simulated using the PSM implementation \citep{PSM2013}. At low redshift ($z < 0.25$), the Compton-$y$ signal is obtained from hydrodynamical simulations of large-scale structure, which include galaxy clusters, diffuse gas, and filamentary structures. At higher redshift ($0.25 < z \leq 1$), galaxy clusters are generated from the Tinker halo mass function \citep{Tinker2008}, assuming \pl\ 2018 cosmological parameters \citep{planck2016-l06}. Cluster masses and redshifts are drawn via Poisson sampling of the mass function, while cluster positions are drawn from a uniform distribution over Galactic coordinates. Masses are restricted to $10^{14}$--$10^{16}\,M_\odot$, and the Compton-y profiles of individual clusters are derived from the universal pressure profile \citep{Arnaud2010}. The resulting $y$-map is scaled across \pl\ frequencies using the non-relativistic thermal SZ SED (Equation~\ref{eq:sed}).

Galactic thermal dust emission is modelled using the \pl\ GNILC dust maps at 353\,GHz in both temperature and polarization \citep{planck2016-XLVIII,planck2016-l04}, extrapolated to other frequencies using a modified blackbody SED with spatially varying temperature and spectral index also provided by GNILC. The synchrotron intensity template is based on the reprocessed Haslam 408\,MHz map \citep{2015MNRAS.451.4311R}, while the polarized synchrotron templates are taken from the \pl\ Commander solution \citep{planck2016-l04}. Both are extrapolated across frequencies assuming a power-law SED with a spatially varying spectral index provided by \cite{Miville2008}. 

All components are co-added to produce sky maps in Stokes $I$, $Q$, and $U$ at 30, 44, 70, 100, 143, 217, and 353\,GHz, and in Stokes $I$ only at 545 and 857\,GHz. The maps are generated at a HEALPix\footnote{\url{https://healpix.jpl.nasa.gov/}} \citep{2005ApJ...622..759G} resolution of $N_{\rm side}=2048$ and convolved with symmetric Gaussian beams with full widths at half maximum (FWHM) corresponding to the \pl\ frequency channels \citep{planck2016-l01}. Instrumental noise is added as Gaussian white noise realizations, with per-pixel RMS values matching the \pl\ channel sensitivities for the \pl-like simulations \citep{planck2016-l01}, and reduced by a factor of $25$ in the low-noise simulations.

\begin{figure}
\centering
\includegraphics[width=0.49\textwidth]{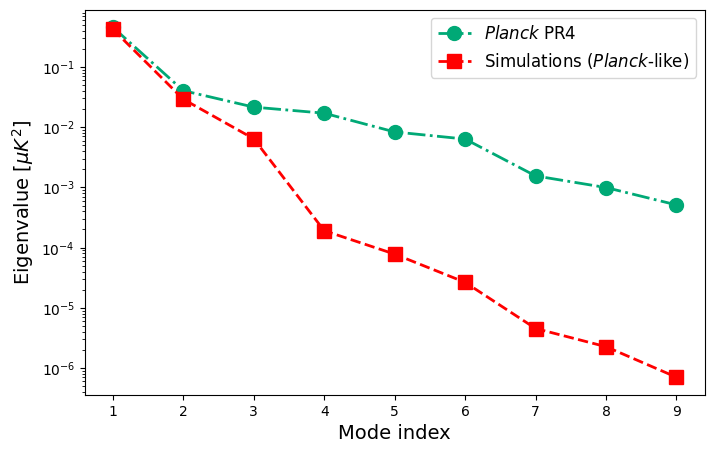}
\includegraphics[width=0.49\textwidth]{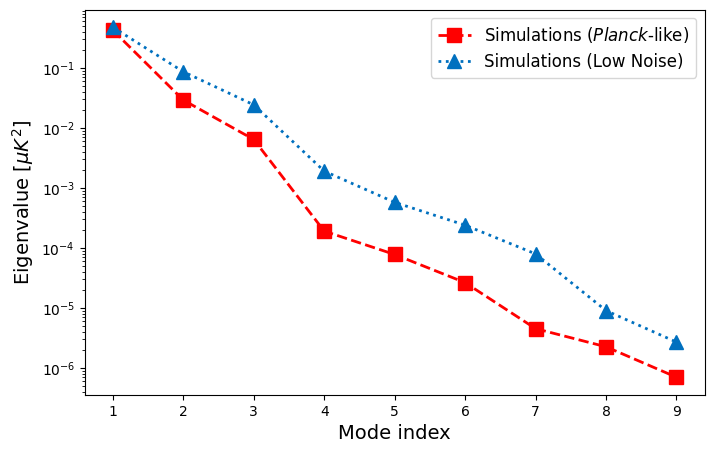}
\caption{Eigenvalues of the correlation matrix $\tP$ (Equation~\ref{eq:pearson}), quantifying the degree of correlation between temperature and polarization channels primarily driven by Galactic foregrounds. Larger eigenvalues indicate stronger correlation, enabling greater Galactic foreground suppression in the $y$-map via the multi-Stokes Hybrid ILC.
\emph{Left}: \pl\ PR4 data (dash-dot green / circles) versus \pl-like simulations (dashed red / squares). The larger eigenvalues in PR4 data indicate stronger intrinsic temperature--polarization correlation of Galactic emission than captured in the simulations, suggesting improved Hybrid ILC performance on real data.
\emph{Right}: \pl-like simulations (dashed red / squares) versus low-noise simulations (blue dotted / triangles). Increased sensitivity enhances the effective temperature--polarization correlation of Galactic foregrounds, as the measured correlation becomes less diluted by instrumental noise.}
\label{fig:eigvals}
\end{figure}

The simulated sky maps exhibit intrinsic temperature--polarization correlations arising mainly from Galactic dust and synchrotron emission, as well as from the CMB, but at a lower level than observed in real PR4 data. This is evident in the left panel of Figure~\ref{fig:eigvals}, which shows that the eigenvalues of the correlation matrix $\tP$ (Equation~\ref{eq:pearson}) are overall smaller for the \pl\ simulations than for the \pl\ PR4 data. This difference reflects the simplified treatment of Galactic foregrounds in the simulations, where single intensity and polarization templates are scaled using simple SED models, as well as the reduced coherence between intensity and polarization templates derived from different data products (e.g. Haslam for synchrotron intensity versus \pl\ Commander for synchrotron polarization). As a result, the performance of the multi-Stokes Hybrid ILC on \pl-like simulations is expected to be less pronounced than on real data. These simulations are therefore primarily intended for a relative comparison between \pl-like and low-noise scenarios (right panel of Figure~\ref{fig:eigvals}), rather than for a direct validation against observations.

%%%%%%%%%%%%%%%%%%%%%%%%%%%%%%%%%%%%%%%%%%%%%%%%%%%%%%%%%%%%%%%%%%%%%%%%%%%%%%%%%%%%%%%%%%%%%%%%%%%%

\section{Analysis and results} \label{sec:results}

The performance of the multi-Stokes Hybrid ILC method, which combines temperature and polarization $Q$ and $U$ frequency maps to enhance Galactic foreground suppression in the reconstructed $y$-map, is assessed in this section. We begin in Section~\ref{sec:NILC_implementation} by describing the needlet-domain implementation of the method and the preprocessing of the input data. The method is then evaluated in Section~\ref{sec:TQU}. It is first applied in Section~\ref{sec:TQU_sims} to both \pl-like simulations and low-noise simulations, the latter constructed by artificially increasing the sensitivity by a factor of 25 to illustrate the improvement achieved at higher sensitivity. The method is subsequently applied to \pl\ PR4 data in Section~\ref{sec:TQU_data}. Results including the additional deprojection of selected sets of CIB moments, using constrained variants of the method, are presented in Section~\ref{sec:deproj_data}.

\subsection{Needlet domain implementation and data preprocessing}
\label{sec:NILC_implementation}

The multi-Stokes Hybrid ILC described in Section~\ref{sec:TP}, as well as its constrained variants (CILC; Section~\ref{sec:deproj}), are implemented in the needlet domain \citep{doi:10.1137/040614359,2008MNRAS.383..539M}, using a set of spherical wavelets that form a tight frame on the sphere. The needlet decomposition closely follows the procedure described in \cite{PR4NILCymap}, but is applied here to both temperature and polarization channels. Each of the nine \pl\ frequency maps in total intensity and the seven frequency maps in both $Q$ and $U$ polarization are decomposed into ten needlet coefficient maps using the set of harmonic-domain needlet windows defined in \citep{PR4NILCymap}. Performing the ILC on needlet coefficients (Needlet ILC, or NILC; \citep{2009A&A...493..835D}), rather than on spherical harmonic coefficients or directly in pixel space, is significantly more effective for reconstructing the thermal SZ signal \citep{PR4NILCymap}. This is because the tSZ signal contains both diffuse and compact features, while its contaminants dominate at different angular scales and sky locations.

The localization properties of needlets in both pixel and harmonic space allow for an adaptive foreground treatment: Galactic foregrounds are preferentially mitigated on large angular scales and at low Galactic latitudes, while CIB contamination and instrumental noise are more efficiently controlled at smaller angular scales and higher latitudes. This spatially and spectrally localized approach enables the implementation of both the Hybrid ILC and the selective CIB moment deprojection strategies within the NILC framework. The ten needlet windows adopted in this work closely follow those used in our previous work for the \pl\ PR4 NILC $y$-map in terms of number and shape (see Section~3.2 and Figure~3 of \citep{PR4NILCymap}).

For the PR4 total-intensity data, the dipole and the frequency-dependent dipole-induced kinematic quadrupole are removed from each of the nine frequency maps \citep{PR4NILCymap}. No analogous preprocessing is applied to the polarization data. A small mask excluding $2\,\%$ of the sky around the central Galactic plane (see Figure~1 of \citep{PR4NILCymap}) is applied to suppress ringing effects from bright Galactic emission during spherical harmonic transforms (SHT). As a result, the reconstructed $y$-maps in this work cover a sky fraction of $f_{\rm sky} = 98\,\%$.

All frequency maps are deconvolved from their native instrumental beams and reconvolved to a common Gaussian beam with a full width at half maximum (FWHM) of $10'$. The PR4 beam transfer functions for polarization differ slightly from those for temperature; this difference is consistently accounted for in the beam deconvolution. The $I$, $Q$, and $U$ frequency maps are first transformed into $T$, $E$, and $B$ spherical harmonic coefficients. Beam reconvolution and needlet decomposition are then applied in harmonic space before transforming back to pixel space to obtain the $I$, $Q$, and $U$ needlet coefficient maps. This procedure is summarized as
\begin{align}
\label{eq:beaming}
x_\nu^{IQU}(p) \;\xrightarrow{\mathrm{SHT}}\; a^{TEB}_{\ell m,\, \nu}
\;\xrightarrow{\times}\;
a^{TEB}_{\ell m,\, \nu}\times\frac{b_\ell^{\,\mathrm{gauss}}}{b_{\ell,\, \nu}^{\,TEB,\,\mathrm{PR4/sims}}}\times h_\ell^j
\;\xrightarrow{\mathrm{SHT^{-1}}}\;
x_\nu^{IQU,\, j}(p)\,,
\end{align}
where $b_\ell^{\,\mathrm{gauss}}$ denotes the spherical harmonic transform of a Gaussian beam with $10'$ FWHM, $b_{\ell,\, \nu}^{\,TEB,\,\mathrm{PR4/sims}}$ is the native beam transfer function for the PR4 data or simulations for temperature or polarization at frequency $\nu$, and $h_\ell^j$ is the $j^{\rm th}$ needlet window in harmonic space.

Subsequent steps, including covariance matrix estimation, weight computation, and ILC map recombination from the needlet coefficient maps $x_\nu^{IQU,\,j}(p)$, are performed following \cite{PR4NILCymap}. 

\subsection{Thermal SZ $y$-map reconstructed from temperature and polarization}
\label{sec:TQU}

Among the nine \pl\ frequency channels ($30$, $44$, $70$, $100$, $143$, $217$, $353$, $545$, and $857$\,GHz), all but the two highest-frequency channels provide polarization observations. We include the Stokes $Q$ and $U$ maps as polarization inputs, as they outperform combinations based on $E$- and $B$-mode maps or on the total polarization amplitude $P$ (see Appendix~\ref{app:QU_choice}). In total, this results in 9 intensity ($T$) channels and 14 polarization channels (7 in $Q$ and 7 in $U$).

In the following, we refer to the standard NILC $y$-map constructed from temperature only as the $T$ $y$-map, and to the multi-Stokes Hybrid NILC map reconstructed from combined $T$, $Q$, and $U$ channels as the $TQU$ $y$-map.

\subsubsection{Forecast for high-sensitivity data from simulations}
\label{sec:TQU_sims}

We assess how the hybrid $TQU$ $y$-map compares to the standard $T$ $y$-map in terms of residual foreground contamination. Two experimental configurations are considered: one with the same sensitivity as \pl, and another with a sensitivity 25 times higher, designed to emulate future high-sensitivity experiments such as \lb. The latter configuration is obtained by using identical simulated sky realizations, but reducing the noise component by a factor of 25 in all channels.

To quantify the relative improvement of the $TQU$ $y$-map with respect to the $T$ $y$-map as a function of multipole, we define the ratio of their angular power spectra, as well as the ratio of the power spectra of their projected Galactic foreground residuals:
\begin{align} \label{eq:cl_diff}
    R_\ell^{\,{\rm tot}} &= \frac{C_\ell^{\,y,\, {\rm tot}\, [TQU]}}{C_\ell^{\,y,\, {\rm tot}\, [T]}}\,, 
    \quad 
    R_\ell^{\,{\rm Gal}} = \frac{C_\ell^{\,y,\, {\rm Gal}\, [TQU]}}{C_\ell^{\,y,\, {\rm Gal}\, [T]}}\,.
\end{align}
Here, $C_\ell^{\,y,\, {\rm tot}\, [TQU]}$ and $C_\ell^{\,y,\, {\rm tot}\, [T]}$ denote the total power spectra of the $TQU$ and $T$ $y$-maps, respectively, while $C_\ell^{\,y,\, {\rm Gal}\, [TQU]}$ and $C_\ell^{\,y,\, {\rm Gal}\, [T]}$ denote the corresponding power spectra of the projected Galactic foreground residuals. The projected residuals are obtained by combining the needlet coefficients of the input Galactic foreground maps in each frequency channel with the corresponding ILC weights derived from the total-sky observation maps using either the multi-Stokes Hybrid NILC or the standard NILC method.

The angular power spectra are computed over $98\,\%$ of the sky (see the mask in Figure~1 of \citep{PR4NILCymap}) using the pseudo-$C_\ell$ estimator \texttt{NaMaster} \cite{2019MNRAS.484.4127A}, which properly accounts for mode coupling due to masking, multipole binning, and beam and pixel window effects.

\begin{figure}[htbp]
    \centering
    \includegraphics[width=\columnwidth]{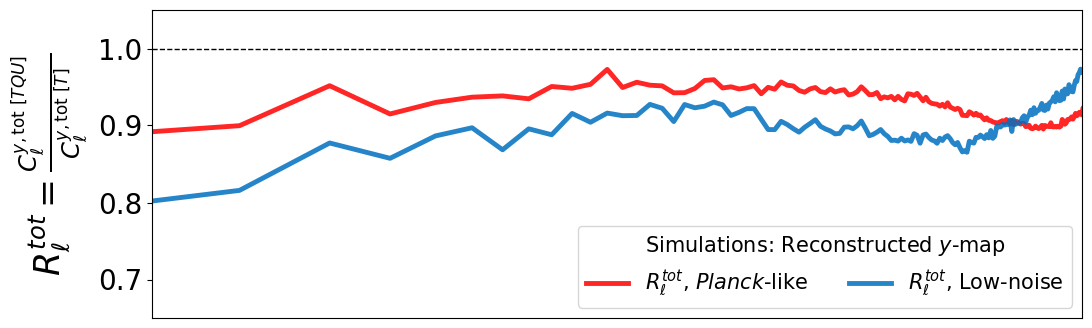}\\[0.5cm]
    \includegraphics[width=\columnwidth]{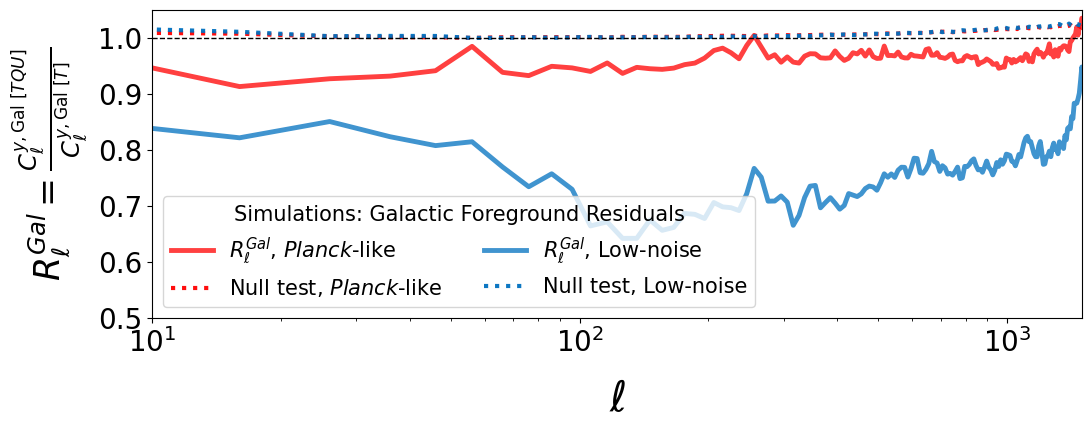}
    \caption{\emph{Top}: Ratio of the total power spectra, $R_\ell^{\,{\rm tot}}$ (Equation~\ref{eq:cl_diff}), between the $TQU$ and $T$ $y$-maps for the \pl-like simulations (solid red) and the low-noise simulations (solid blue). 
\emph{Bottom}: Ratio of the Galactic foreground residual power spectra, $R_\ell^{\,{\rm Gal}}$ (Equation~\ref{eq:cl_diff}), between the $TQU$ and $T$ $y$-maps for the same simulations. 
Dotted lines show null tests in which Galactic components are removed from the polarization $Q$ and $U$ channel maps prior to the $y$-map reconstruction.  These results demonstrate that the suppression of Galactic foreground residuals in the $TQU$ $y$-map occurs only when temperature and polarization foreground components are correlated, and that the level of suppression increases with improved sensitivity.}
    \label{fig:sims_T_vs_TQU_spectra}
\end{figure}

Figure~\ref{fig:sims_T_vs_TQU_spectra} shows the ratios $R_\ell^{\,{\rm tot}}$ (top panel) and $R_\ell^{\,{\rm Gal}}$ (bottom panel), computed with multipole bins of $\Delta \ell = 10$, for the \pl-like (red) and low-noise (blue) simulations. 
In both configurations, the $TQU$ $y$-map exhibits lower power than the $T$ $y$-map across the multipoles, as indicated by ratios significantly below unity. The reduction is even more pronounced in the low-noise case (blue curves), confirming our expectations: reducing instrumental noise increases the effective correlation between temperature and polarization channels (as illustrated in the right panel of Figure~\ref{fig:eigvals}), thereby enhancing the variance suppression predicted by Equation~\eqref{eq:varratio}.

The overall reduction of total power in the $TQU$ $y$-map is primarily driven by improved mitigation of Galactic foreground residuals, as shown in the bottom panel of Figure~\ref{fig:sims_T_vs_TQU_spectra}. The average percentage reductions across multipoles are summarized in Table~\ref{tab:power_reduction_sims}. For \pl-like simulations, Galactic residuals decrease by about $6\,\%$ over low multipoles $\ell \in [10,150]$ and by about $3\,\%$ over the full multipole range. These values should not be interpreted as representative of the actual \pl\ data, for which we expect larger values, since the simulations do not fully reproduce the level of temperature--polarization correlation observed in PR4 data (left panel of Figure~\ref{fig:eigvals}). More relevant is the relative improvement obtained when increasing experimental sensitivity.  In the low-noise configuration, corresponding to a 25-fold sensitivity enhancement comparable to \lb, the reduction of residual Galactic contamination in the $TQU$ $y$-map relative to the $T$-only $y$-map is further enhanced, reaching about $23$--$26\,\%$ across multipoles (Table~\ref{tab:power_reduction_sims}).

As a null test, Galactic components are removed from the polarization channels before reconstructing the $TQU$ $y$-map with the multi-Stokes Hybrid NILC method. In this case, the ratio of Galactic foreground residuals between the $TQU$ and $T$ $y$-maps is consistent with unity (dotted lines in the bottom panel of Figure~\ref{fig:sims_T_vs_TQU_spectra}), indicating no significant difference between the two reconstructions in the absence of temperature--polarization correlations, independently of experimental sensitivity. This confirms that the observed reduction in Galactic contamination arises from correlated foreground components present in both temperature and polarization channels, which the multi-Stokes Hybrid NILC method exploits.

\begin{table}
\centering
\caption{Average percentage reduction in power of the $TQU$ $y$-map relative to the $T$ $y$-map, defined as $\left(R_\ell^{\,{\rm tot}} - 1\right)\times 100$, and of their projected Galactic foreground residuals, defined as $\left(R_\ell^{\,{\rm Gal}} - 1\right)\times 100$. Results are shown for \pl-like and low-noise simulations, averaged over the multipole ranges $\ell \in [10,150]$ and $\ell \in [10,1500]$.}
\label{tab:power_reduction_sims}
\begin{tabular}{llcc}
\toprule
Simulation & Quantity & $\ell \in [10,150]$ & $\ell \in [10,1500]$ \\
\midrule
\multirow{2}{*}{\pl-like} 
  & Total $y$-map power & $-5.8\,\%$ & $-7.7\,\%$ \\
  & Galactic residual power & $-5.8\,\%$ & $-3.3\,\%$ \\
\midrule
\multirow{2}{*}{Low-noise} 
  & Total $y$-map power & $-10.9\,\%$ & $-9.5\,\%$ \\
  & Galactic residual power & $-25.7\,\%$ & $-23.2\,\%$ \\
\bottomrule
\end{tabular}
\end{table}

Artificial chance correlations between the thermal SZ signal and foregrounds at low multipoles can induce signal loss, known as ILC bias \cite{2009A&A...493..835D}. The ILC bias scales with the number of channels and inversely with the number of modes used in both harmonic and spatial domains for NILC localization. Since the $TQU$ Hybrid NILC reconstruction uses more channels than the standard $T$-only NILC, the ILC bias could in principle increase if no mitigation strategy is applied. Such strategies include increasing the effective number of modes by reducing spatial localization or excluding central modes of the localization domain (in pixel or harmonic space) when estimating the covariance matrix, thereby decorrelating the NILC weights from the data (see e.g. \cite{PhysRevD.109.063530}). Quantitatively, we find that the additional signal loss in the $TQU$ $y$-map remains below 1\,\% of the total power of the $T$ $y$-map across multipoles when adopting identical localization schemes. This contribution is therefore negligible compared to the overall power reduction associated with Galactic foreground mitigation observed in Figure~\ref{fig:sims_T_vs_TQU_spectra}.

These results demonstrate that including polarization channels in NILC improves Galactic foreground mitigation in the reconstructed thermal SZ $y$-map over a wide range of angular scales for \pl-level sensitivity. For future high-sensitivity experiments such as \lb\ \cite{LiteBIRD:2024dbi}, where instrumental noise becomes subdominant, the suppression of Galactic residuals is expected to be substantially stronger.

\begin{figure}[htbp]
    \centering
    \includegraphics[width=0.75\columnwidth]{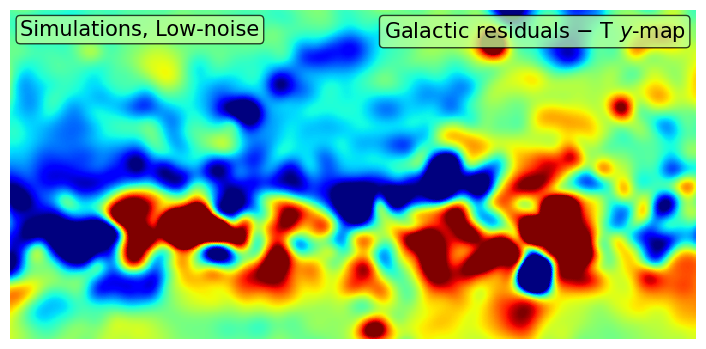}
    \includegraphics[width=0.75\columnwidth]{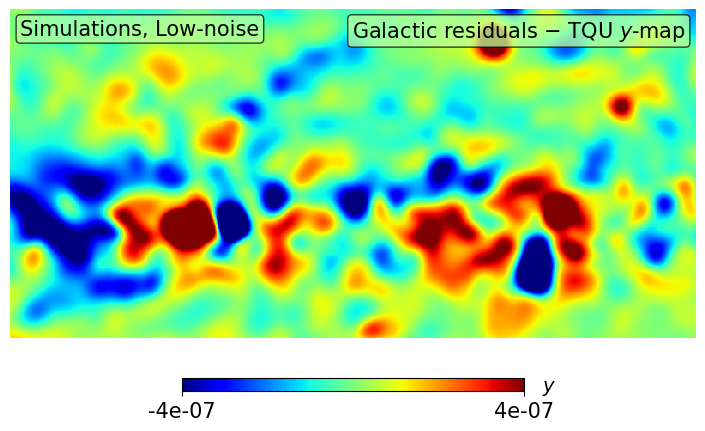}
    \caption{$50\deg \times 24\deg$ strip around the Galactic plane, centred at Galactic coordinates ${(\ell,b)=(125\deg,5\deg)}$, showing Galactic residual maps (smoothed to $90'$) for the $T$ (top) and $TQU$ (bottom) $y$-map reconstructions in the low-noise simulation. The RMS values are $2.82\times10^{-7}$ and $2.02\times10^{-7}$, respectively, corresponding to about $30\,\%$ reduction in residual Galactic contamination for the $TQU$ $y$-map in this region.}
    \label{fig:sims_T_vs_TQU_gal_res_strip}
\end{figure}

To further demonstrate the reduction of Galactic foreground residuals in the $TQU$ $y$-map, we examine specific sky regions. Galactic residual maps are constructed by applying either the hybrid or standard NILC weights to the needlet coefficients of the input Galactic foreground components only, rather than to those of the full simulated sky. Figure~\ref{fig:sims_T_vs_TQU_gal_res_strip} shows a $50\deg \times 24\deg$ strip around the Galactic plane for the low-noise simulation. The root mean square (RMS) value of the Galactic residuals in this region is significantly lower in the $TQU$ reconstruction compared to the $T$ reconstruction, decreasing from $2.82\times10^{-7}$ to $2.02\times10^{-7}$, corresponding to a reduction of approximately $30\,\%$. This visual inspection confirms the enhanced mitigation of Galactic foreground contamination in the reconstructed $y$-map when polarization information is included.

\begin{figure}[htbp]
    \centering
    \includegraphics[width=0.55\columnwidth]{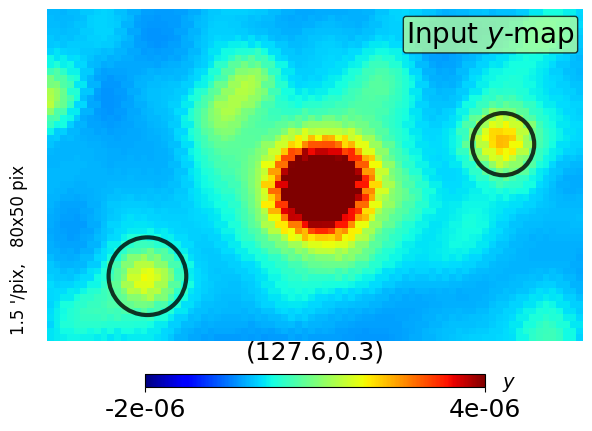}\\[0.3cm]
    \includegraphics[width=0.47\columnwidth]{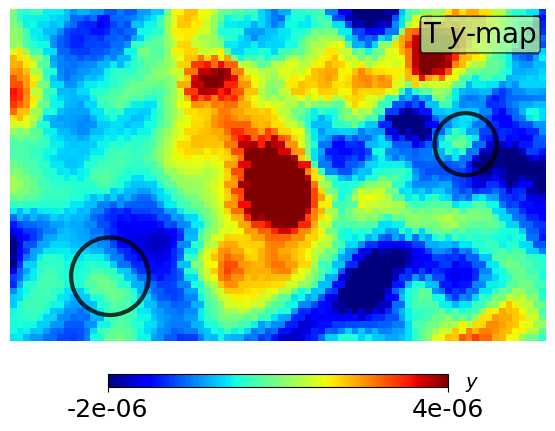}
    \includegraphics[width=0.47\columnwidth]{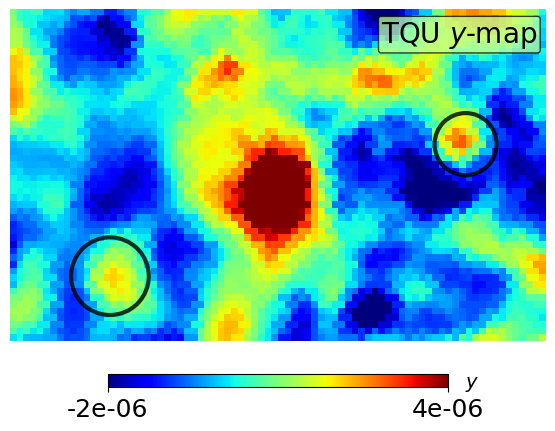}
    \includegraphics[width=0.47\columnwidth]{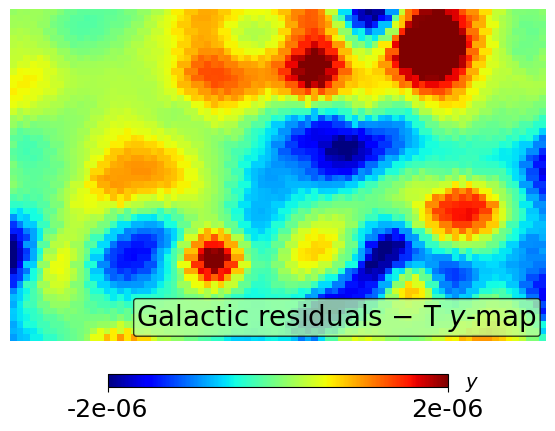}
    \includegraphics[width=0.47\columnwidth]{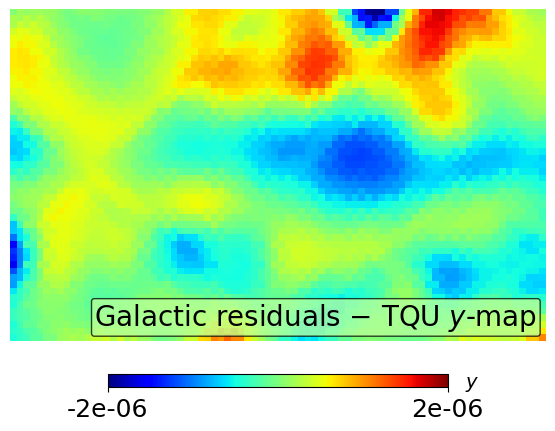}
    \caption{\emph{Top}: Input $y$-map showing three galaxy clusters in the Galactic plane region. \emph{Middle row}: $T$ $y$-map reconstruction (left) and $TQU$ $y$-map reconstruction (right) for the low-noise simulation. \emph{Bottom row}: Corresponding Galactic residual maps for the $T$ (left) and $TQU$ (right) reconstructions. Circled regions highlight fainter nearby clusters visible in the input map that are recovered in the $TQU$ reconstruction but not in the $T$ reconstruction.}
    \label{fig:sims_T_vs_TQU_gal_res_clusters}
\end{figure}

Figure~\ref{fig:sims_T_vs_TQU_gal_res_clusters} illustrates three galaxy clusters located near the Galactic plane, where foreground contamination is expected to be strongest, together with their reconstruction in the low-noise simulation. Compared to the standard $T$ $y$-map reconstruction (middle left), the $TQU$ $y$-map reconstruction (middle right) exhibits a clearer and better-resolved thermal SZ signal. The corresponding Galactic residual maps (bottom panels) show that the surrounding foreground contamination is substantially reduced in the $TQU$ case. Two fainter clusters visible in the input $y$-map (circled regions in the top panel) are recovered in the $TQU$ reconstruction but remain obscured in the $T$ reconstruction. Inspection of the residual maps indicates that this improved recovery is primarily driven by the lower level of Galactic foreground contamination in the vicinity of these compact sources.

\subsubsection{Application to \pl\ PR4 data}
\label{sec:TQU_data}

We now apply the multi-Stokes Hybrid NILC to the \pl\ PR4 data described in Section~\ref{sec:data}, producing a PR4 $TQU$ $y$-map and comparing it to the public PR4 NILC $T$ $y$-map released by \cite{PR4NILCymap}. The processing of the PR4 $TQU$ $y$-map closely follows that of the PR4 NILC $y$-map presented in \cite{PR4NILCymap}, with the only modification being the inclusion of the additional polarization channels described in Section~\ref{sec:TP}. 

\begin{figure}[htbp]
    \centering
    \includegraphics[width=0.32\columnwidth]{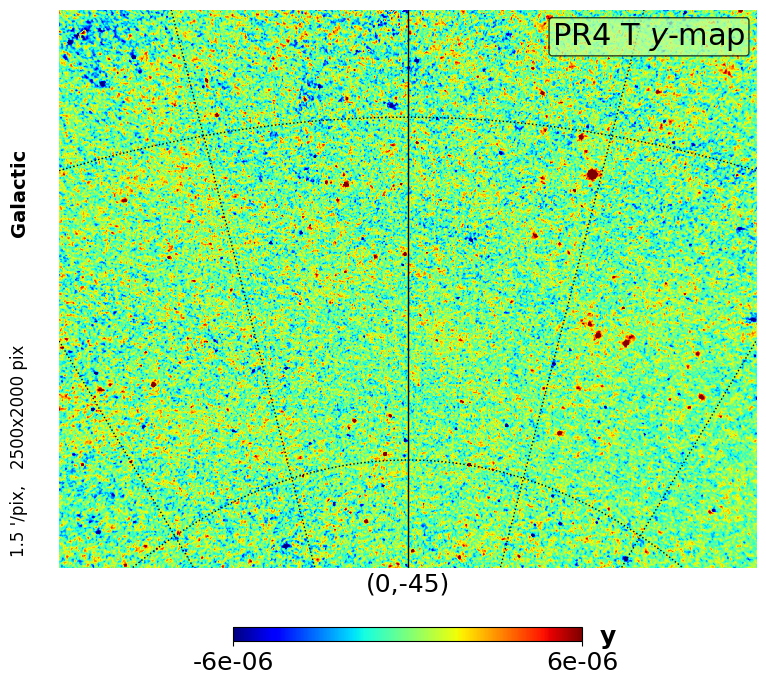}
    \includegraphics[width=0.32\columnwidth]{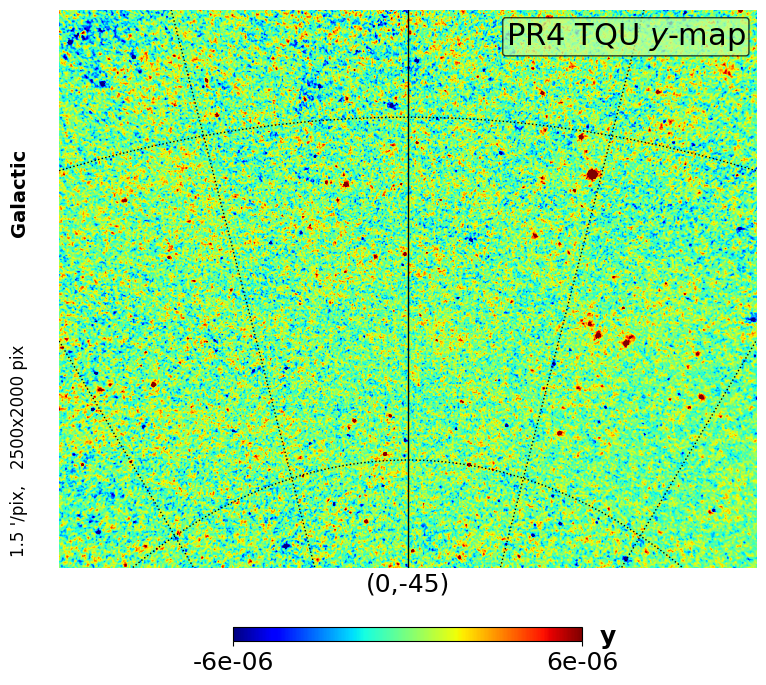}
    \includegraphics[width=0.32\columnwidth]{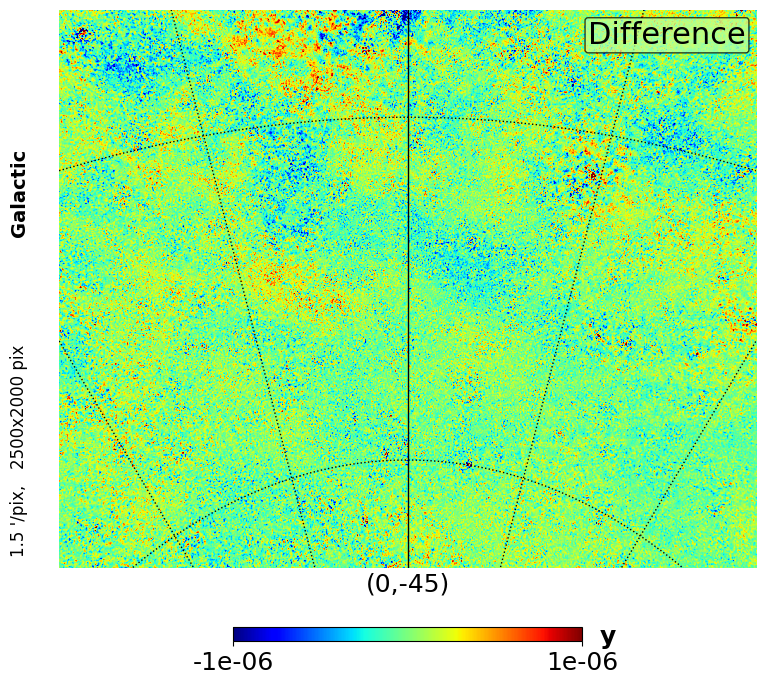}
      \caption{A $62.5\deg\times 50\deg$ region of the sky, centred at Galactic coordinates $(\ell,b)=(0\deg,-45\deg)$, showing the PR4 $T$ $y$-map (left, \cite{PR4NILCymap}), the PR4 $TQU$ $y$-map (middle, this work), and the difference between the two maps (right; colour-scale range reduced by a factor of 6 to highlight low-amplitude features). The two reconstructions appear visually very similar, demonstrating consistent recovery of the thermal SZ signal from galaxy clusters, while their difference highlights diffuse large-scale features consistent with differing levels of residual foreground contamination.}
    \label{fig:pr4_region}
\end{figure}

Figure~\ref{fig:pr4_region} shows the PR4 NILC $T$ $y$-map \cite{PR4NILCymap} in the left panel and the PR4 $TQU$ $y$-map (this work) in the middle panel, displayed over the same $62.5\deg \times 50\deg$ region of the sky centred at Galactic coordinates $(\ell,b)=(0\deg,-45\deg)$. The two $y$-maps exhibit consistent reconstruction of the thermal SZ signal from galaxy clusters, visible as red spots in the maps. The difference between the two $y$-maps, shown in the right panel, exhibits diffuse large-scale features consistent with residual Galactic foreground contamination that is more prominent in the $T$ $y$-map, as discussed below.

\begin{figure}[htbp]
    \centering
    \includegraphics[width=0.49\columnwidth]{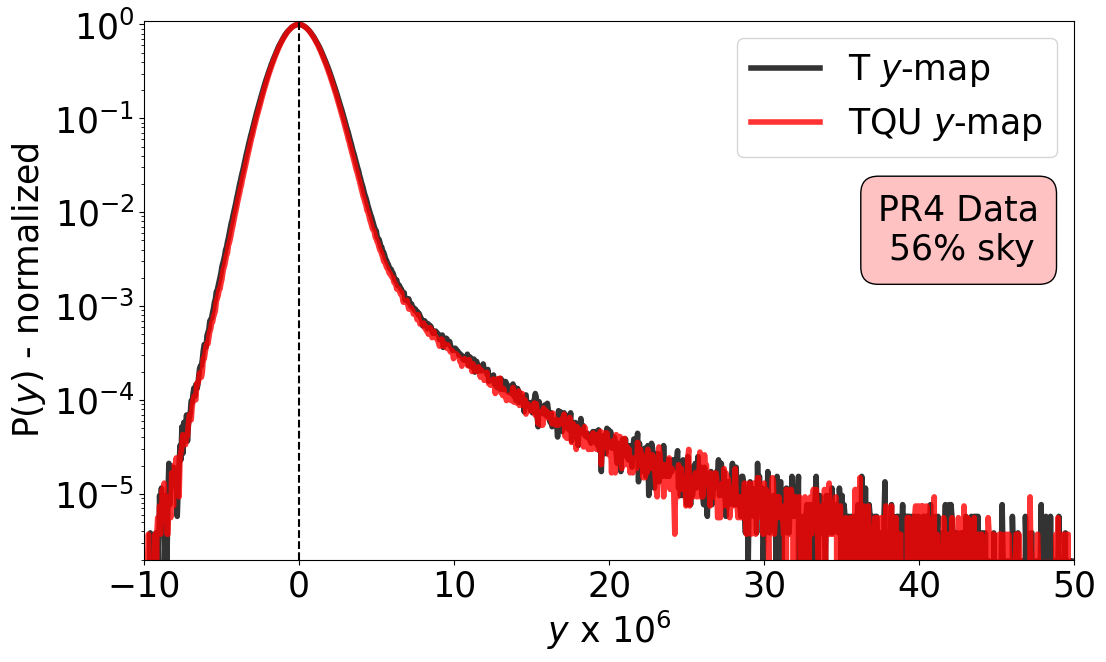}
    \includegraphics[width=0.49\columnwidth]{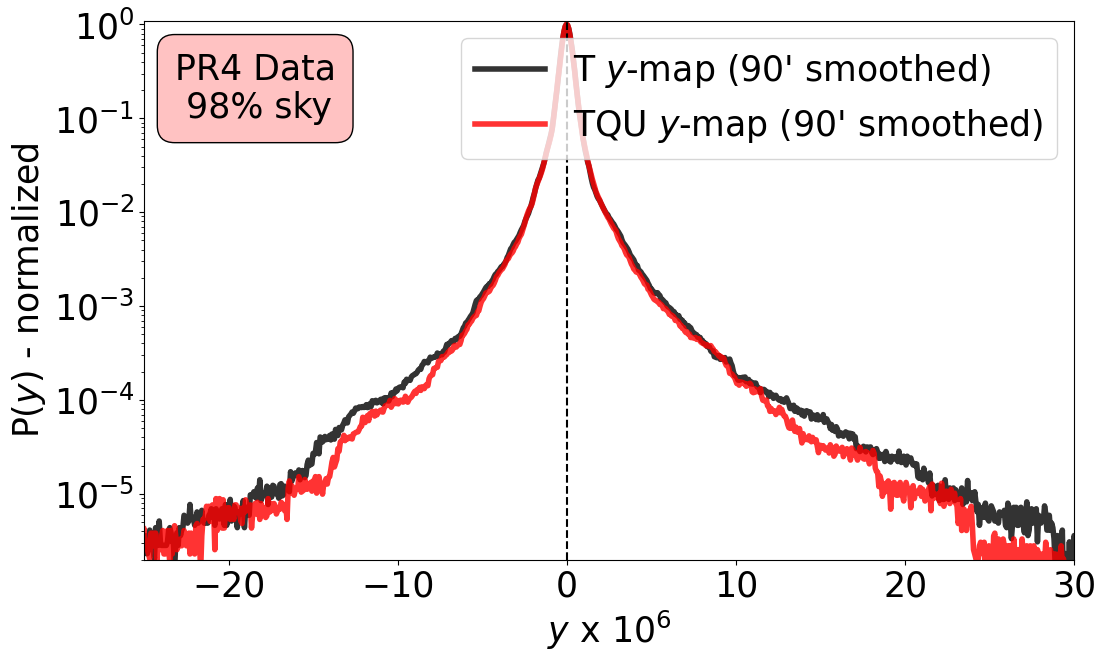}
	 \caption{\emph{Left}: Normalized one-point probability distribution function (1-PDF) of the PR4 NILC $y$-maps obtained from the $T$ (black, \cite{PR4NILCymap}) and $TQU$ (red, this work) reconstructions, computed over $56\,\%$ of the sky, showing consistent recovery of the skewed tail of the thermal SZ signal. \emph{Right}: Normalized 1-PDFs of the two $y$-maps, computed over $98\,\%$ of the sky after smoothing to $90'$. The enlarged sky fraction and the additional smoothing are intended to emphasize the contribution of diffuse, large-scale Galactic residual contamination, which is visibly reduced in the $TQU$ $y$-map.}
    \label{fig:pdf_pr4}
\end{figure}

The normalized one-point probability distribution functions (1-PDFs) of both $y$-maps, computed over $f_{\rm sky}=56\,\%$ of the sky\footnote{Using the union of the $f_{\rm sky}=60\,\%$ Galactic mask from \cite{PR4NILCymap} and a point-source mask constructed from \cite{2006MNRAS.370.2047L,planck2014-a35}.} (see the mask in Figure~1 of \cite{PR4NILCymap}) and shown in the left panel of Figure~\ref{fig:pdf_pr4}, are also consistent, recovering the characteristic skewed tail of the thermal SZ effect and thereby corroborating the robustness of the SZ signal reconstruction. Despite this apparent agreement in thermal SZ recovery, the two $y$-maps differ significantly in their level of residual Galactic foreground contamination, as we illustrate below.

The right panel of Figure~\ref{fig:pdf_pr4} shows the normalized 1-PDFs of the PR4 $TQU$ and $T$ $y$-maps, computed over $98\,\%$ of the sky after smoothing to $90'$, to emphasize the contribution from diffuse, large-scale Galactic residual contamination. Differences between the two reconstructions are most pronounced in bright Galactic regions, which contribute to the tails of the distribution away from zero. The 1-PDF of the PR4 $TQU$ $y$-map (red) is visibly narrower than that of the PR4 $T$ $y$-map (black), particularly in the tails. This provides a first quantitative indication that the PR4 $TQU$ $y$-map exhibits reduced large-scale Galactic residual contamination compared to the public PR4 NILC $T$ $y$-map.

\begin{figure}[htbp]
    \centering
    \includegraphics[width=0.85\columnwidth]{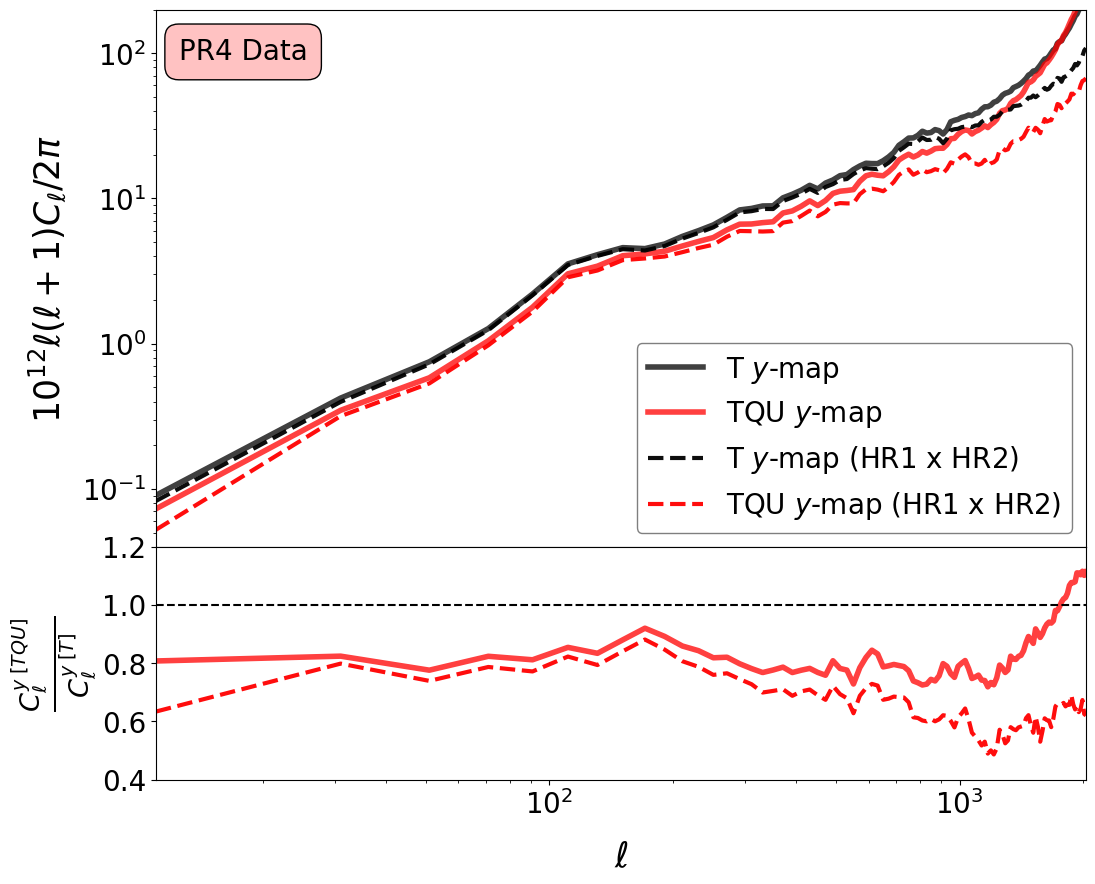}
\caption{\emph{Top}: Angular power spectra of the PR4 $T$ $y$-map (black, \cite{PR4NILCymap}) and the PR4 $TQU$ $y$-map (red, this work), computed over $f_{\rm sky}=98\,\%$ of the sky, where diffuse Galactic residuals are more prominent. Solid lines show the auto-power spectra (prior to noise-bias correction), while dashed lines show the cross-power spectra between the HR1 and HR2 $y$-maps, which remove the noise bias. \emph{Bottom}: Ratio of the PR4 $TQU$ to $T$ $y$-map power spectra, illustrating the reduction in overall power across multipoles for the $TQU$ reconstruction.}
    \label{fig:cl_pr4}
\end{figure}

The angular power spectra of the PR4 $TQU$ and $T$ $y$-maps, computed over $f_{\rm sky}=98\,\%$ of the sky where diffuse Galactic contamination is more prominent, are shown in the top panel of Figure~\ref{fig:cl_pr4}. Both the auto-spectra (solid lines) and the cross-spectra between PR4 half-ring 1 (HR1) and half-ring 2 (HR2) $y$-maps (dashed lines) are displayed. The half-ring $y$-maps are obtained by applying the full-data NILC weights to the HR1 and HR2 data splits, whose noise contributions are expected to be mostly uncorrelated. The cross-spectrum of these $y$-maps therefore largely removes the noise bias.

The ratio of the PR4 $TQU$ to $T$ $y$-map power spectra is shown in the bottom panel of Figure~\ref{fig:cl_pr4}, and the corresponding average percentage of power reduction over multipoles is reported in Table~\ref{tab:power_reduction_pr4}. The overall power of the PR4 $TQU$ $y$-map is significantly reduced compared to that of the PR4 $T$ $y$-map across angular scales, with an average reduction of approximately $20\,\%$. When considering the cross-spectra, which suppress the noise bias and therefore emphasize foreground contribution, the reduction becomes even more pronounced (approximately $35\,\%$). We note that a smaller level of improvement, at the level of $\sim 6\,\%$, is obtained when applying the same Galactic mask used for the left panel of Figure~\ref{fig:pdf_pr4}, corresponding to $f_{\rm sky}=56\,\%$. This is consistent with the fact that this mask excludes the regions most affected by Galactic foreground residuals, where the improvement from the $TQU$ reconstruction is most significant.

The reduction in power is considerably larger in the PR4 data (Table~\ref{tab:power_reduction_pr4}) than in the \pl-like simulations (Table~\ref{tab:power_reduction_sims}). This difference arises because current simulations do not fully capture the level of Galactic foreground temperature--polarization correlations present in the real data, as discussed in Section~\ref{sec:sims} and illustrated in the left panel of Figure~\ref{fig:eigvals}.

\begin{figure}[htbp]
    \centering
    \includegraphics[width=\columnwidth]{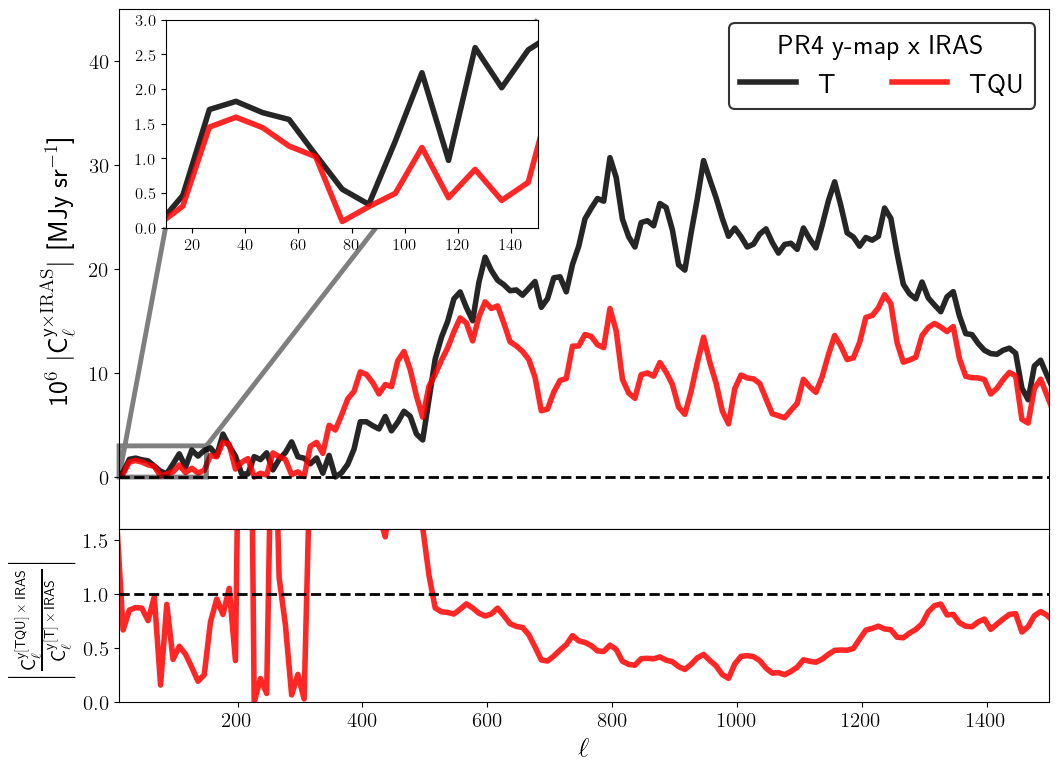}
 \caption{\emph{Top}: Cross-power spectra between the PR4 $y$-maps and the IRAS $100\,\micron$ map, shown in absolute value, for the $T$ (black) and $TQU$ (red) reconstructions. The inset shows a zoom of low multipoles $\ell \leq 150$. 
\emph{Bottom}: Ratio of the $TQU$ to $T$ $y \times {\rm IRAS}$ cross-spectra. The PR4 $TQU$ $y$-map (red, this work) exhibits a lower overall correlation with the IRAS dust tracer than the public PR4 $T$ $y$-map (black, \cite{PR4NILCymap}).}
    \label{fig:iras_cross_pr4}
\end{figure}

\begin{table}[htbp]
\centering
\caption{Average percentage reduction in power of the PR4 $TQU$ $y$-map (this work) relative to the PR4 $T$ $y$-map \cite{PR4NILCymap} over the multipole ranges $\ell \in [10,150]$ and $\ell \in [10,1500]$. The first row reports the average of $100\times (C_\ell^{\,y[TQU]} - C_\ell^{\,y[T]})/C_\ell^{\,y[T]}$ for the auto-power spectra, while the second row gives the same quantity for the half-ring cross-spectra. The third row shows the average of $100\times (C_\ell^{\,y[TQU] \times {\rm IRAS}} - C_\ell^{\,y[T] \times {\rm IRAS}})/C_\ell^{\,y[T] \times {\rm IRAS}}$, corresponding to the percentage reduction in residual Galactic foreground contamination estimated from the cross-correlation of the reconstructed $y$-maps with the IRAS $100\,\micron$ map.}
\label{tab:power_reduction_pr4}
\begin{tabular}{lccc}
\toprule
Quantity & $\ell \in [10,150]$ & $\ell \in [10,1500]$ \\
\midrule
$y\times y$ (full) & $-17.8\,\%$ & $-20.7\,\%$ \\
$y \times y$ (HR1 $\times$ HR2) & $-21.7\,\%$  & $-34.5\,\%$ \\
$y \times {\rm IRAS}$ (Galactic residuals) & $-45.4\,\%$ & $-40.8\,\%$\\
\bottomrule
\end{tabular}
\end{table}

\begin{figure}[htbp]
    \centering
    \includegraphics[width=0.8\columnwidth]{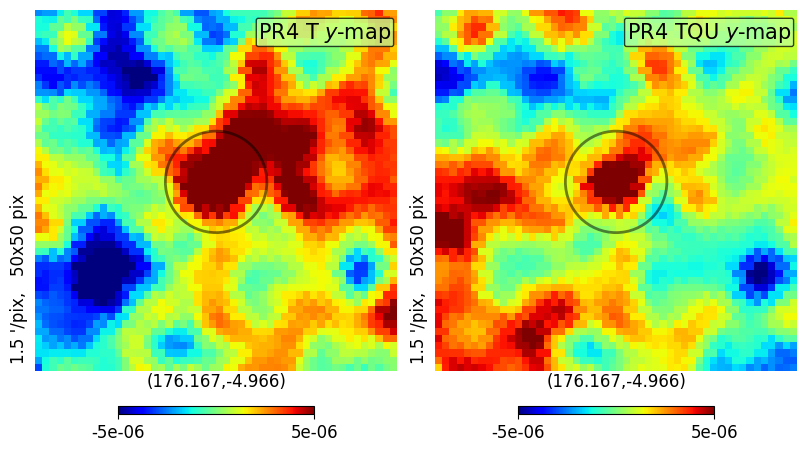}
    \caption{Galaxy cluster CIZA J0516.9+2925 \cite{2002ApJ...580..774E,2007ApJ...662..224K} located in the Galactic plane region, shown in the PR4 NILC $T$ $y$-map (left, \cite{PR4NILCymap}) and the PR4 $TQU$ $y$-map (right, this work). The cluster is better resolved in the $TQU$ reconstruction due to reduced residual Galactic contamination. The RMS values in the displayed region are $2.76\times10^{-6}$ and $1.95\times10^{-6}$ for the $T$ and $TQU$ $y$-maps, respectively, corresponding to a $\sim 30\,\%$ reduction in RMS.}
    \label{fig:pr4_cluster}
\end{figure}

To verify that the reduction in power in the PR4 $TQU$ $y$-map is due to lower Galactic foreground residuals, we compute the cross-power spectra of the PR4 $T$ and $TQU$ $y$-maps with the IRAS $100\,\micron$ map \cite{2005ApJS..157..302M}, an external tracer of Galactic thermal dust emission. The top panel of Figure~\ref{fig:iras_cross_pr4} shows these $y \times {\rm IRAS}$ cross-spectra, while the bottom panel displays the ratio of the $TQU$ to $T$ $y \times {\rm IRAS}$ cross-spectra. A clear reduction in cross-correlation power is observed for the $TQU$ $y$-map compared to the $T$ $y$-map across multipoles. The low multipoles, where Galactic foregrounds dominate, are highlighted in the inset for clarity. The average reduction in thermal dust residuals from $T$ to $TQU$ is $\sim 40$--$45\,\%$ (Table~\ref{tab:power_reduction_pr4}, third row), providing strong evidence of lower Galactic contamination in the PR4 $TQU$ $y$-map. This demonstrates, in the case of real \pl\ PR4 data, that including polarization channels in NILC improves the mitigation of Galactic foregrounds in the $y$-map, in agreement with both our analytical predictions (Section~\ref{sec:TP}) and simulation results (Section~\ref{sec:TQU_sims}).

Reduced diffuse Galactic foreground contamination in the $y$-map should, in principle, enable improved recovery of galaxy clusters near the Galactic plane. Figure~\ref{fig:pr4_cluster} shows one such cluster, CIZA J0516.9+2925 \cite{2002ApJ...580..774E,2007ApJ...662..224K}, located at Galactic coordinates $(\ell,b)=(176.2\deg,-5\deg)$. The cluster is better resolved in the PR4 $TQU$ $y$-map than in the PR4 $T$ $y$-map. The RMS value in the $1.25\deg \times 1.25\deg$ region centred on the cluster is lower in the PR4 $TQU$ $y$-map by approximately $30\,\%$, consistent with reduced diffuse Galactic foreground contamination and with the behaviour observed in simulations (Figure~\ref{fig:sims_T_vs_TQU_gal_res_clusters}).

As expected for \pl\ data, we do not identify many such clusters near the Galactic plane that are clearly better resolved in the $TQU$ reconstruction. However, simulations presented in Section~\ref{sec:TQU_sims} show that this multi-Stokes cleaning strategy becomes more effective for low-noise experiments. We therefore anticipate that future high-sensitivity CMB experiments beyond \pl\ will reveal a larger number of better-resolved clusters at low Galactic latitudes using $TQU$ $y$-map reconstructions.

\subsection{Thermal SZ $y$-map reconstructed with selective CIB moment deprojection}
\label{sec:deproj_data}

Unlike Galactic foregrounds, which primarily contaminate the $y$-maps at large angular scales and are strongly polarized, the CIB contributes mainly at small angular scales and is essentially unpolarized. Consequently, the multi-Stokes NILC approach discussed in Section~\ref{sec:TP} has little direct impact on residual CIB contamination, and a different mitigation strategy is required. In particular, CIB contamination can be reduced by deprojecting its spectral moments using a Constrained (needlet) ILC (CILC), as described in Section~\ref{sec:deproj}.

Here, we apply CILC to all \pl\ PR4 $I$, $Q$, and $U$ frequency maps within the multi-Stokes framework. The nulling constraints on CIB moments are imposed only on the $I$ channels. Including the $Q$ and $U$ polarization channels does not directly affect the CIB deprojection, since the CIB is unpolarized; however, by reducing Galactic contamination, they can indirectly facilitate improved CIB minimization within the ILC. In practice, we find that the resulting changes in residual CIB contamination are negligible compared to those induced by the choice of CIB deprojection itself.

The aim of this analysis is to identify the optimal combination of moments to deproject and the needlet bands over which to apply the deprojection, such that residual CIB contamination is minimized while the noise penalty in the final $y$-map remains as small as possible. Similar approaches were explored in \cite{NILC-McCarthyHill}, and a comparison is provided in Appendix~\ref{app:cib_deproj_comparison}. Here, however, we test previously unexplored combinations of moments and needlet bands, as summarized in Table~\ref{tab:cilc}, to determine the most effective configuration. 

\begin{figure}[htbp]
    \centering   
    \includegraphics[width=0.97\columnwidth]{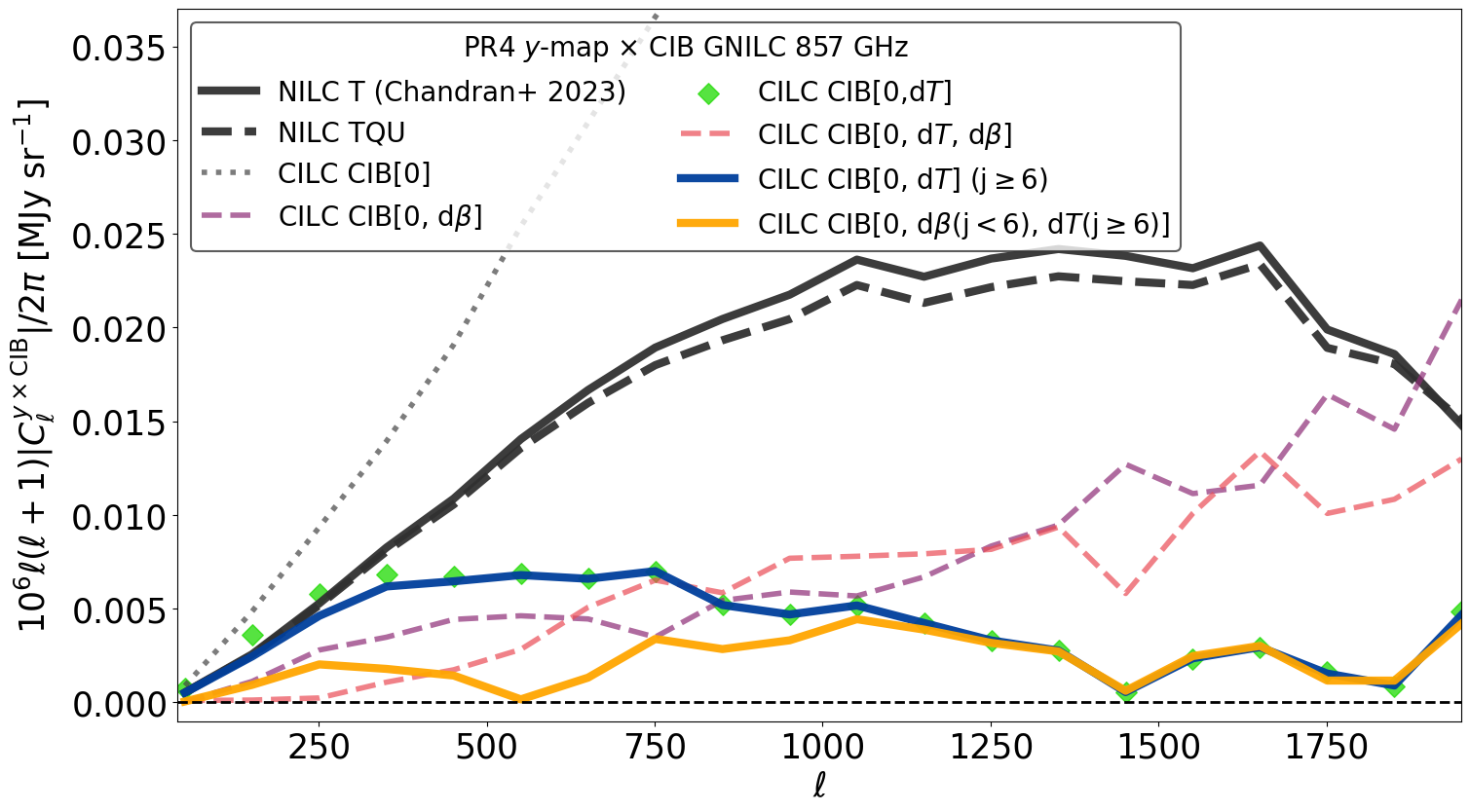}
    \includegraphics[width=0.97\columnwidth]{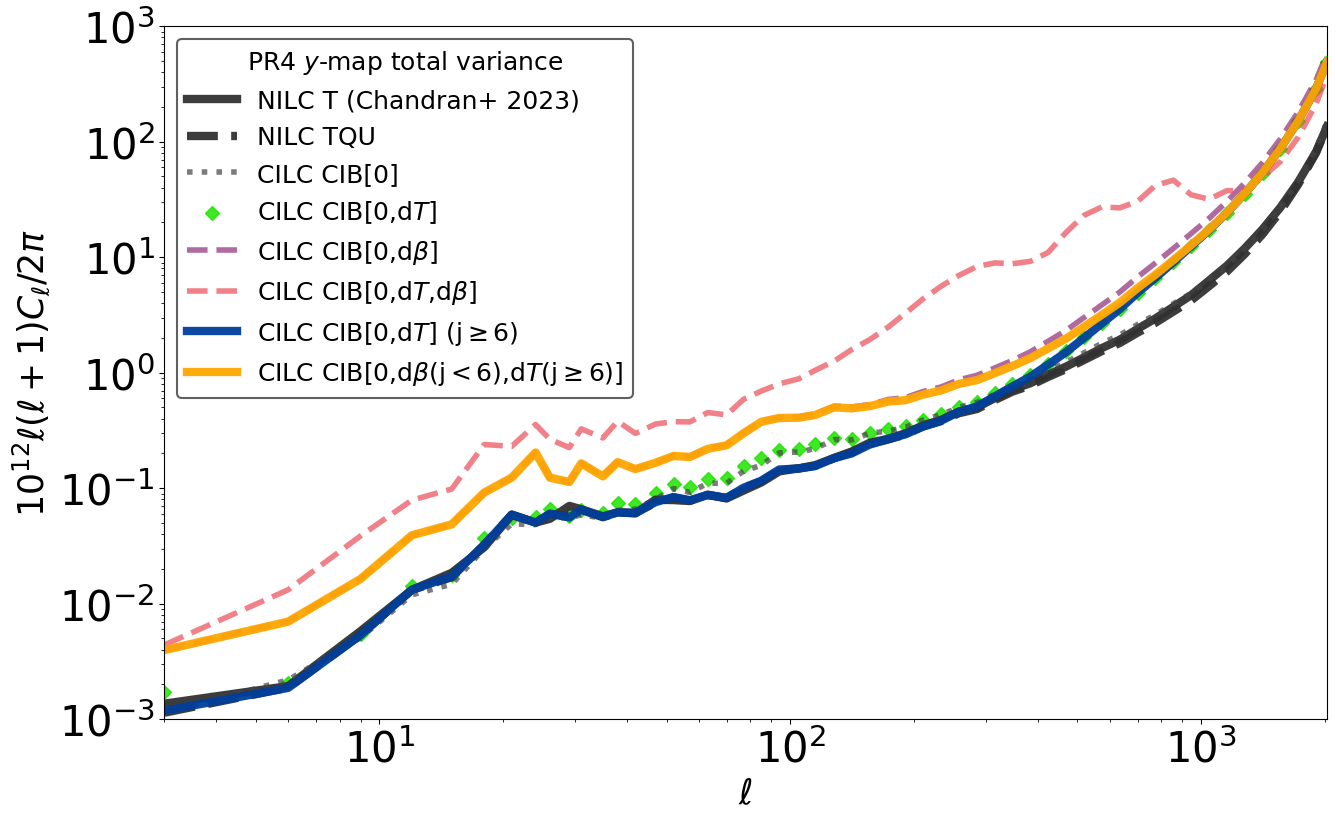}
	\caption{\emph{Top}: The $y \times \text{CIB}$ cross-power spectra between the reconstructed PR4 $y$-maps and the \pl\ GNILC CIB map at $857$\,GHz over $f_{\rm sky}=50\,\%$ of the sky, highlighting the level of residual CIB contamination in the different PR4 $y$-maps. See Table~\ref{tab:cilc} for the denomination of the various CILC maps in the legend, corresponding to the specific sets of deprojected CIB moments and the needlet bands (indexed by $j$) in which they are deprojected. \emph{Bottom}: The $y \times y$ auto-power spectra of each reconstructed PR4 $y$-map, highlighting the associated noise (including unconstrained foreground) penalty for each CILC deprojection configuration.}
    \label{fig:cib_deproj_cl}
\end{figure}

To assess residual CIB contamination in the resulting $y$-maps, we compare their $y \times \text{CIB}$ cross-power spectra using the \pl\ GNILC CIB $857$\,GHz map \cite{planck2016-XLVIII} as the CIB tracer, following \cite{PR4NILCymap}. While this cross-spectrum may contain a contribution from the intrinsic astrophysical tSZ--CIB correlation, that contribution is expected to be approximately common to all reconstructed $y$-maps since the tSZ signal is preserved by construction. Therefore, differences in the measured cross-spectra primarily trace the relative level of residual CIB contamination among the different $y$-map reconstructions. The results are shown in the top panel of Figure~\ref{fig:cib_deproj_cl}, while the bottom panel displays the corresponding $y$-map auto-power spectra, illustrating the associated noise (including unconstrained foreground) penalty for each CIB deprojection configuration.

Looking at the two panels of Figure~\ref{fig:cib_deproj_cl}, several conclusions can be drawn regarding the basic deprojection combinations. First, the reduced CIB cross-power observed for the PR4 NILC $TQU$ $y$-map (dashed black) relative to the PR4 NILC $T$ $y$-map (solid black) likely reflects lower residual Galactic foreground contamination—correlated with residual thermal dust in the GNILC map—rather than a genuine reduction in CIB contamination. That said, the inclusion of polarization channels reduces Galactic residuals and may indirectly grant the ILC additional freedom to further minimize CIB variance in the intensity channels.

Deprojecting only the zeroth-order moment of the CIB (CILC$[0]$, gray dotted line) increases residuals relative to blind variance minimization with NILC (black lines). This reflects both the limitations of an imperfect mean CIB SED model and the variance penalty arising from unconstrained higher-order CIB moments. Incorporating the derivative of the SED with respect to spectral index, thereby deprojecting the first-order moment in $\beta$ (CILC$[0,d\beta]$, purple dashed line), significantly reduces CIB residuals at low multipoles and remains below the no-deprojection NILC cases across all multipoles. However, this improvement comes at the cost of a substantial increase in total variance at all scales (bottom panel).

 By contrast, deprojecting the first-order temperature moment (CILC$[0,dT]$, green diamonds) provides strong suppression of CIB residuals at high multipoles with only a minimal variance penalty at low $\ell$, making it the most promising of the basic configurations. Although its CIB residuals are slightly higher at $\ell \lesssim 250$ than those of CILC$[0,d\beta]$---a regime where CIB is not the dominant contaminant---it achieves the most favourable overall balance between CIB suppression and noise increase. Deprojecting all three moments (CILC$[0,d\beta,dT]$, salmon-pink dashed line) yields only marginal improvement over CILC$[0,d\beta]$ at high multipoles, while incurring a substantial variance penalty across all multipoles (bottom panel).

These results indicate that deprojecting only the zeroth-order CIB moment is ineffective, and that deprojecting all three zeroth- and first-order moments is not justified given the large variance penalty. Among the basic combinations, CILC$[0,dT]$ offers the best compromise between CIB suppression and total variance increase. Since CILC$[0,dT]$ performs best at high multipoles, whereas CILC$[0,d\beta]$ is more effective at low multipoles, we construct scale-dependent hybrid configurations in which different moments are deprojected in different needlet bands to combine the advantages of both approaches.

Through empirical exploration, we identify two hybrid configurations that offer the best compromise between CIB suppression and variance increase. The first, CILC$[0,dT]\,(j\geq 6)$ (blue solid line), applies the $0$ and $dT$ deprojections only in needlet bands 6--10 ($\ell \gtrsim 700$). This configuration achieves excellent small-scale CIB suppression while maintaining a low-$\ell$ variance nearly indistinguishable from the no-deprojection NILC cases, and therefore provides the best overall balance. 

The second configuration, CILC$[0,d\beta(j<6),dT(j\geq 6)]$ (yellow solid line), applies zeroth-order moment deprojection at all scales, $d\beta$ deprojection at large scales (needlet bands 1--5), and $dT$ deprojection at small scales (needlet bands 6--10). This strategy yields the lowest overall CIB contamination but at the cost of increased large-scale variance relative to the previous configuration.

For most applications of an all-sky thermal SZ $y$-map requiring minimal CIB contamination, CILC$[0,dT]\,(j\geq 6)$ appears optimal. The more aggressive configuration, CILC$[0,d\beta(j<6),dT(j\geq 6)]$, may be preferable in cases where the lowest possible CIB contamination is required even on large angular scales, though the resulting map is significantly noisier. All-sky $y$-maps corresponding to the configurations discussed here are shown in Figure~\ref{fig:cilc_maps} in Appendix~\ref{app:cib_deproj_comparison}. 

In principle, this optimization could be automated through a cost function balancing CIB suppression against overall signal-to-noise ratio, as in \cite{Carones2024}. However, current sensitivity levels do not warrant such refinement for cosmological analyses. Such an approach will become increasingly relevant for future high-sensitivity experiments, where additional components can be deprojected with a smaller relative noise penalty.

%%%%%%%%%%%%%%%%%%%%%%%%%%%%%%%%%%%%%%%%%%%%%%%%%%%%%%%%%%%%%%%%%%%%%%%%%%%%%%%%%%%%%%%%%%%%%%%%%%%%

\section{Conclusions} \label{sec:conclusions}

We have presented two complementary extensions to NILC for thermal SZ $y$-map reconstruction: (i) a multi-Stokes Hybrid NILC that combines temperature and polarization channels to suppress Galactic foreground contamination, and (ii) a strategy of selective CIB moment deprojection within CILC to control extragalactic contamination at small angular scales. The multi-Stokes Hybrid NILC leverages intrinsic $TE/TB$ correlations of dust and synchrotron emission to reduce their contribution while preserving the unpolarized thermal SZ signal. The selective CIB moment deprojection balances bias reduction against variance increase by applying scale-dependent nulling constraints to specific CIB moments, while leaving the others to blind variance minimization.

Using \pl\ PR4 data, the hybrid $TQU$ $y$-map reconstruction shows a $\sim 40\,\%$ reduction in correlation with the IRAS $100\,\micron$ dust tracer relative to the standard PR4 NILC $T$-only $y$-map (Table~\ref{tab:power_reduction_pr4}; Figure~\ref{fig:iras_cross_pr4}), indicating significantly lower residual Galactic contamination. Although this improvement does not substantially increase the usable sky fraction for PR4, since the reduction is spatially non-uniform and the brightest Galactic regions remain challenging, it yields a cleaner, lower-variance $y$-map. The reduction is stronger than in \pl-like simulations, reflecting the simplified temperature--polarization coherence in the simulated Galactic templates. Nevertheless, these simulations serve as a proof of concept for future surveys: in low-noise simulations with 25 times better sensitivity than \pl, the gain in Galactic foreground suppression achieved by the $TQU$ $y$-map relative to the $T$-only $y$-map is significantly enhanced, demonstrating that the multi-Stokes Hybrid ILC approach becomes increasingly effective as instrumental noise decreases, as expected for upcoming experiments such as SO and \lb.

We further show that moment deprojection within CILC enables control of CIB-induced biases in the \pl\ PR4 $y$-map at high multipoles, but that indiscriminate nulling of CIB moments leads to a substantial increase in overall variance (Figure~\ref{fig:cib_deproj_cl}). Among the basic configurations considered, CILC$[0,dT]$---which deprojects only the first-order temperature moment while leaving spectral index moments to blind variance minimization---provides the best overall trade-off for the chosen pivot values derived from the mean redshift of \pl\ clusters. This configuration strongly reduces small-scale CIB contamination while keeping the total variance close to that of the minimum-variance NILC map. Building on this, the scale-dependent configuration CILC$[0,dT],(j\geq 6)$ yields the most balanced $y$-map in terms of residual CIB contamination and noise for \pl\ data, whereas the more aggressive CILC$[0,d\beta(j<6),dT(j\geq 6)]$ configuration further suppresses CIB residuals but at the cost of significantly increased noise.

While the multi-Stokes Hybrid NILC achieves a substantial reduction of Galactic foreground contamination, its performance on \pl\ PR4 data remains limited by instrumental noise. The \pl-like simulations, although they do not fully capture the complexity of Galactic dust and temperature--polarization correlations, provide a controlled framework to assess how the cleaning efficiency scales with sensitivity. They indicate that the full potential of the multi-Stokes ILC approach will be realized with future low-noise experiments such as SO and \lb.

More generally, the multi-Stokes Hybrid ILC framework is applicable to other component-separation problems in which polarization provides additional discriminating power. Notable examples include the separation of spectrally degenerate CIB and thermal dust emission, as well as kinetic SZ and CMB anisotropies, the latter benefiting from the fact that the CMB is polarized while the kinetic SZ effect is not. These applications are left for future work.

The PR4 CILC$[0,dT]\,(j\geq 6)$ $TQU$ $y$-map, together with corresponding half-ring splits, will be made publicly available at \url{https://doi.org/10.5281/zenodo.18876142}.\footnote{The $y$-maps with other CIB deprojection configurations can be provided upon request.}

%%%%%%%%%%%%%%%%%%%%%%%%%%%%%%%%%%%%%%%%%%%%%%%%%%%%%%%%%%%%%%%%%%%%%%%%%%%%%%%%%%%%%%%%%%%%%%%%%%%%

\acknowledgments
We thank the anonymous referee for constructive comments that helped improve the quality and clarity of this paper, and Fiona McCarthy for useful discussions.
JC acknowledges financial support from the \textit{Concepci\'on Arenal Programme} of the Universidad de Cantabria.
The authors acknowledge support by the Spanish Ministry of Science and Innovation (MCIN) and the Agencia Estatal de Investigación (AEI) through the project grants PID2022-139223OB-C21, PID2022-140670NA-I00, and PID2025-173927NB-I00. 
This work also received support from the RadioForegroundsPlus Project HORIZONCL4-2023-SPACE-01, GA 101135036.
We also acknowledge support from the Universidad de Cantabria and the Consejer\'ia de Educaci\'on, Formaci\'on profesional y Universidades del Gobierno de Cantabria via the "Actividad estructural para el desarrollo de la investigaci\'on del Instituto de F\'isica de Cantabria" project. 
Some of the presented results are based on observations obtained with \pl,\footnote{http://www.esa.int/Planck} an ESA science mission with instruments and contributions directly funded by ESA Member States, NASA, and Canada.
This work made use of \texttt{Python} packages like \texttt{pymaster} \cite{2019MNRAS.484.4127A}, \texttt{astropy} \cite{2022ApJ...935..167A}, \texttt{scipy} \cite{2020SciPy-NMeth}, and \texttt{matplotlib} \cite{Hunter:2007}. Some of the results in this paper have been derived using the \texttt{healpy} \cite{2019JOSS....4.1298Z} and \textit{HEALPix} \cite{2005ApJ...622..759G} packages. We acknowledge the use of the PSM, developed by the Component Separation Working Group (WG2) of the \pl\ Collaboration.
We acknowledge Santander Supercomputacion support group at the University of Cantabria who provided access to the supercomputer Altamira Supercomputer at the Institute of Physics of Cantabria (IFCA-CSIC), member of the Spanish Supercomputing Network, for performing simulations/analyses.
%

%%%%%%%%%%%%%%%%%%%%%%%%%%%%%%%%%%%%%%%%%%%%%%%%%%%%%%%%%%%%%%%%%%%%%%%%%%%%%%%%%%%%%%%%%%%%%%%%%%%%

% Bibliography

\bibliographystyle{JHEP}
\bibliography{references,Planck_bib}

%%%%%%%%%%%%%%%%% APPENDICES %%%%%%%%%%%%%%%%%%%%%

\appendix

\section{Choice of polarization fields for multi-Stokes Hybrid ILC}
\label{app:QU_choice}

When combining temperature and polarization channels within the multi-Stokes Hybrid NILC framework, several options for the polarization fields are possible: the Stokes $Q$ and $U$ frequency maps (hereafter $TQU$), the $E$- and $B$-mode frequency maps (hereafter $TEB$), or the polarization amplitude maps $P=\sqrt{Q^{2}+U^{2}}$ (hereafter $TP$).

We perform $y$-map reconstruction for each of these three cases using \pl-like simulations, low-noise simulations, and \pl\ PR4 data, and compare the level of Galactic residuals in the final $y$-maps in order to determine the optimal choice. The $TP$ combination presents practical challenges for PR4 data. Computing $P$ requires squaring and summing $Q$ and $U$, which amplifies noise systematics such as scan-direction striping present in \pl\ polarization maps. Mitigating this effect requires smoothing the input $Q$ and $U$ maps prior to constructing $P$, which suppresses small-scale Galactic structure and partially degrades the information content. Although this limitation is not present in simulations, we nevertheless explore the $TP$ case for completeness and for future reference in the context of low-systematics data.

\begin{figure}[htbp]
    \centering
    \includegraphics[width=\columnwidth]{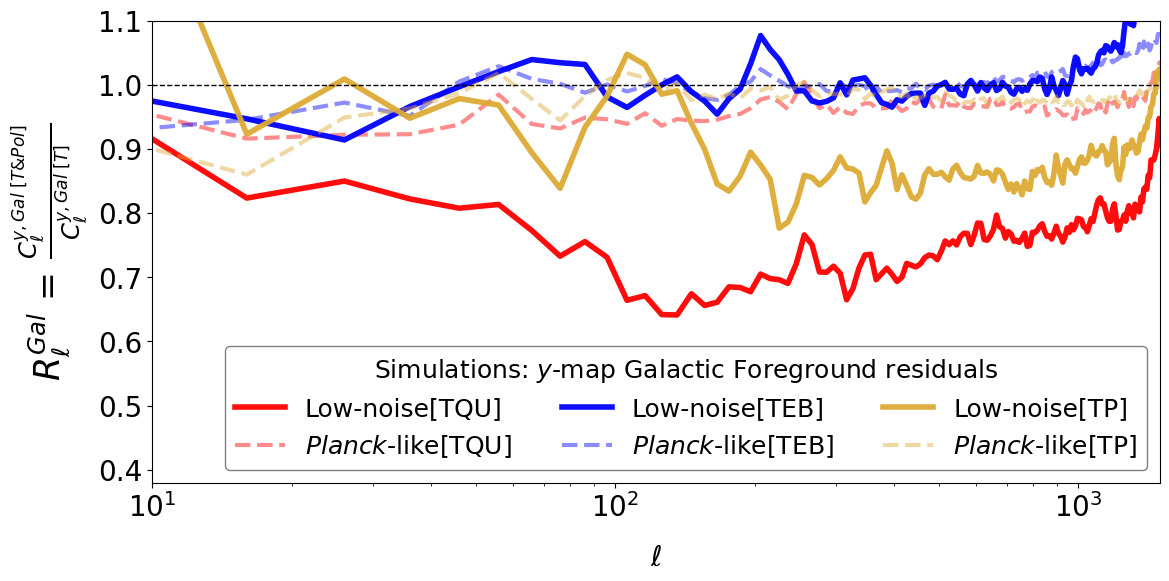}
    \caption{Ratio of Galactic foreground residual power spectra, $R_\ell^{\,{\rm Gal}}$ (Equation~\ref{eq:cl_diff}), between the multi-Stokes ($TQU$, $TEB$, or $TP$) and $T$-only $y$-maps for \pl-like (dashed lines) and low-noise (solid lines) simulations. The $TQU$ implementation (red) consistently yields the lowest residual Galactic contamination, irrespective of instrumental sensitivity.}
    \label{fig:sims_gal_t_pol}
\end{figure}

Figure~\ref{fig:sims_gal_t_pol} shows the ratio of residual Galactic foreground power between the multi-Stokes ($TQU$, $TEB$, or $TP$) and $T$-only $y$-maps for \pl-like (dashed lines) and low-noise (solid lines) simulations. The projected Galactic residuals are obtained by applying the ILC weights—derived from the total maps—to the needlet coefficients of the input Galactic components. In both simulation sets, the $TQU$ combination yields the lowest residual Galactic contamination, followed by $TP$, while $TEB$ provides only marginal improvement over the $T$-only reconstruction. The difference between the implementations becomes more pronounced in the low-noise simulations, demonstrating that the cleaning efficiency of the multi-Stokes approach improves with sensitivity. In the low-noise case, the RMS of the projected Galactic residuals over $f_{\rm sky}=98\,\%$ is reduced by $14.1\,\%$ for $TQU$, compared to $5.7\,\%$ for $TP$ and only $0.1\,\%$ for $TEB$, relative to residuals in the $T$-only $y$-map.

\begin{figure}[htbp]
    \centering
    \includegraphics[width=\columnwidth]{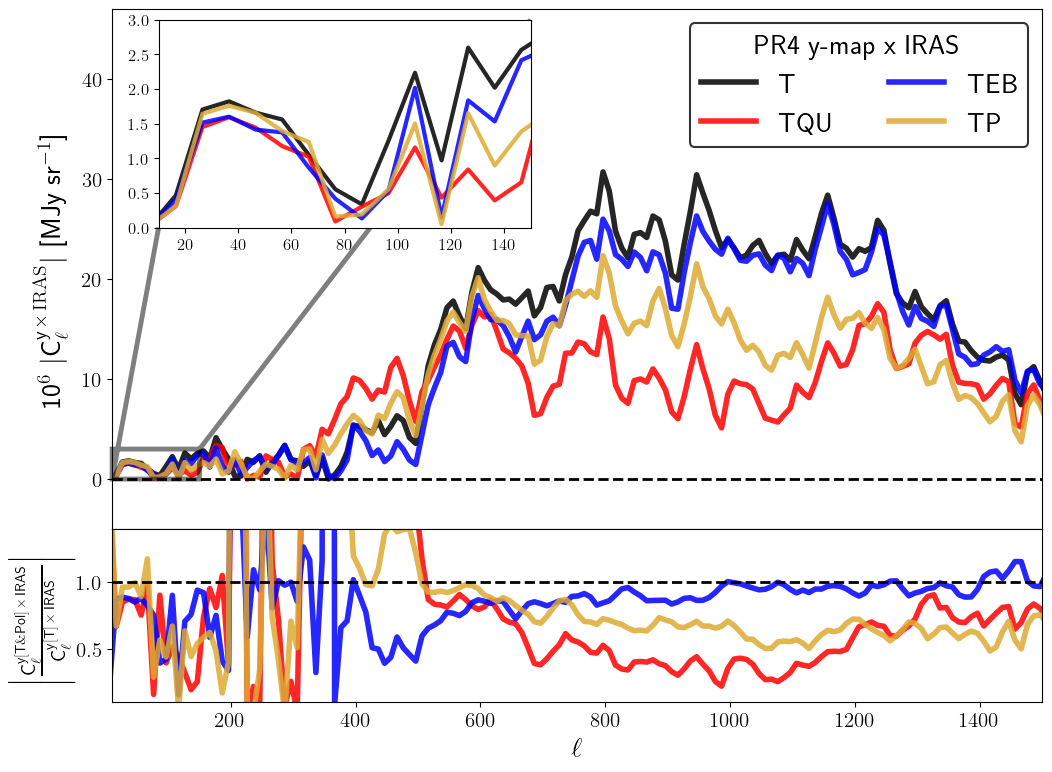}
	 \caption{\emph{Upper panel:} Cross-power spectra of the \pl\ PR4 $y$-maps, reconstructed using different combinations of temperature and polarization channels, with the IRAS $100\,\micron$ map tracing thermal dust emission over $f_{\rm sky}=98\,\%$ of the sky. 
    \emph{Lower panel:} Ratio of the multi-Stokes to $T$-only $y \times \text{IRAS}$ cross-spectra. The $TQU$ reconstruction (red) exhibits the lowest correlation with the dust tracer, followed by $TP$ (yellow) and $TEB$ (blue), relative to the $T$-only case (black).}
    \label{fig:pr4_y_x_iras}
\end{figure}

As discussed in Section~\ref{sec:TQU_data}, current \pl-like simulations use simplified Galactic templates that likely underestimate the true temperature--polarization correlations present in the data. Nevertheless, the overall trends seen in simulations are consistent with PR4 results. Figure~\ref{fig:pr4_y_x_iras} shows the cross-power spectra of the PR4 $y$-maps with the IRAS $100\,\micron$ dust tracer over $f_{\rm sky}=98\,\%$ of the sky. The $TQU$ implementation exhibits the lowest residual dust correlation, followed by $TP$ and then $TEB$. Averaged over $10 < \ell < 150$, the improvement relative to the $T$-only reconstruction is $45\,\%$ for $TQU$, $31\,\%$ for $TP$, and $22\,\%$ for $TEB$; over the full multipole range, the corresponding values are $38\,\%$, $25\,\%$, and $4\,\%$. The qualitative hierarchy matches that observed in simulations.

In both simulations and data, the improvement from $T$ to $TEB$ implementation is minimal compared to other combinations. The better performance of $TQU$ relative to $TEB$ can be understood physically. Thermal dust and synchrotron emission exhibit strong spatial correlations between intensity and polarization patterns. These correlations are directly encoded in the Stokes $Q$ and $U$ maps, which preserve locality and thus maintain pixel-level correspondence with temperature maps. By contrast, the transformation from $Q/U$ to $E/B$ involves non-local operations, redistributing information across the sky. For a spatially-localized, needlet-based ILC implementation, this loss of locality reduces the ability of the algorithm to exploit temperature--polarization correlations for foreground suppression. Incorporating $Q$ and $U$ rather than $E$ and $B$ frequency maps therefore enables more efficient removal of polarized Galactic contaminants in the reconstructed thermal SZ $y$-map.

A conceptual question arises as to whether it is appropriate to linearly combine the scalar temperature field with the spin-2 Stokes $Q$ and $U$ fields. In the present application, this poses no fundamental issue. The thermal SZ signal is intrinsically unpolarized, and the constraint structure of the multi-Stokes ILC ensures zero response of the reconstructed signal to polarization channels. The $Q$ and $U$ maps thus act purely as auxiliary templates correlated with foreground residuals. Although $Q$ and $U$ depend on the chosen polarization basis, the statistical properties of the foreground and noise fields are invariant under rotations of that basis. Consequently, no systematic bias is introduced into the recovered $y$-parameter, and the spinor nature of the polarization field does not affect the reconstruction of an unpolarized signal.

\section{Comparison with other contemporary $y$-maps}
\label{app:cib_deproj_comparison}

\begin{figure}
    \centering
    \includegraphics[width=0.92\textwidth]{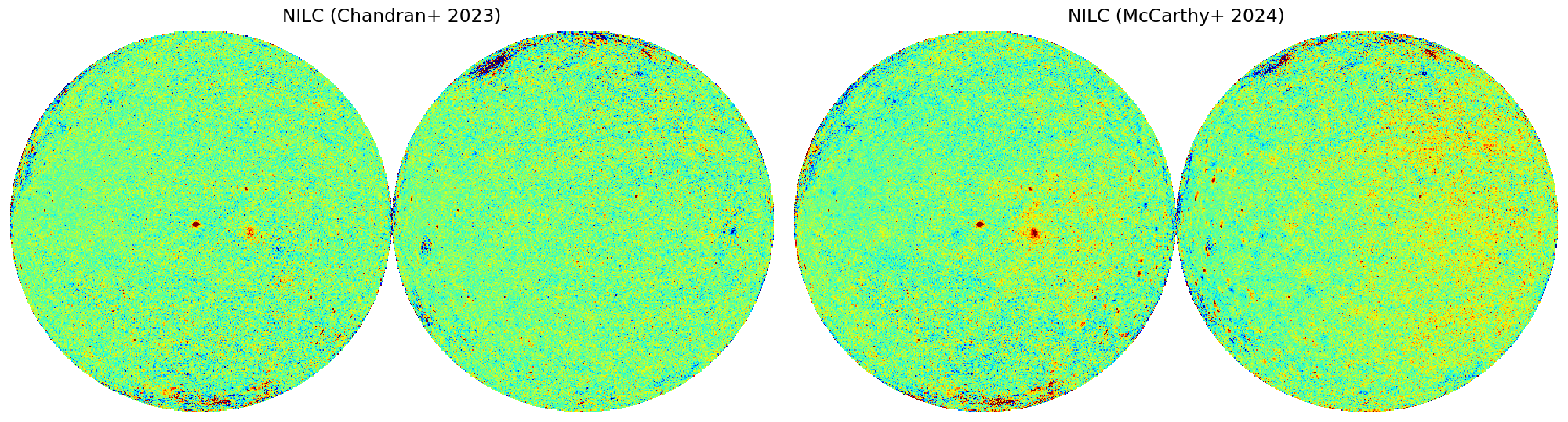}\\
    \includegraphics[width=0.92\textwidth]{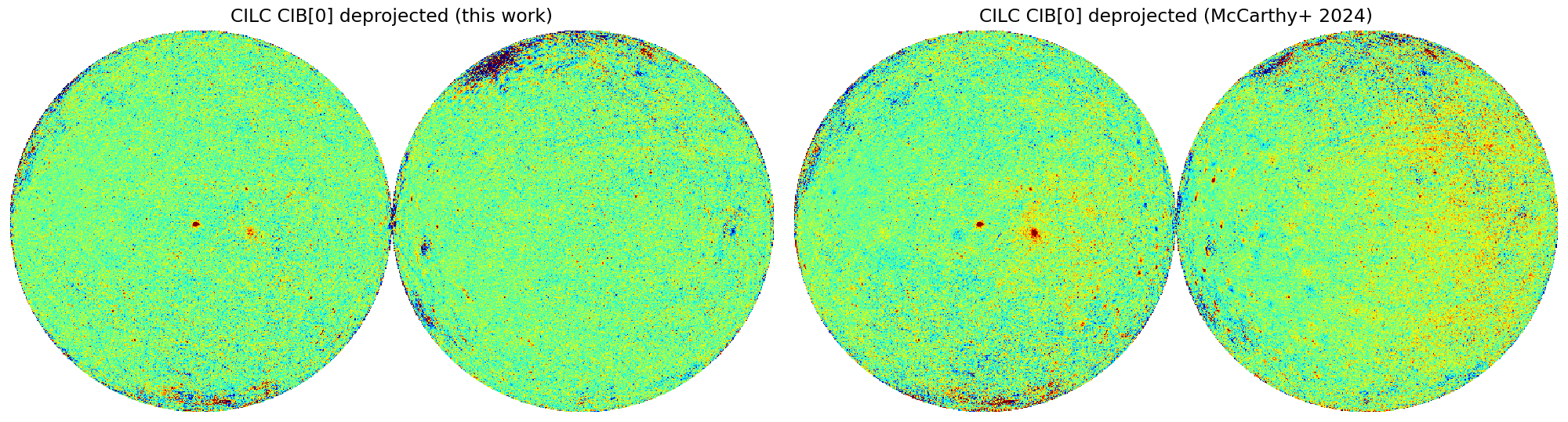}\\
    \includegraphics[width=0.92\textwidth]{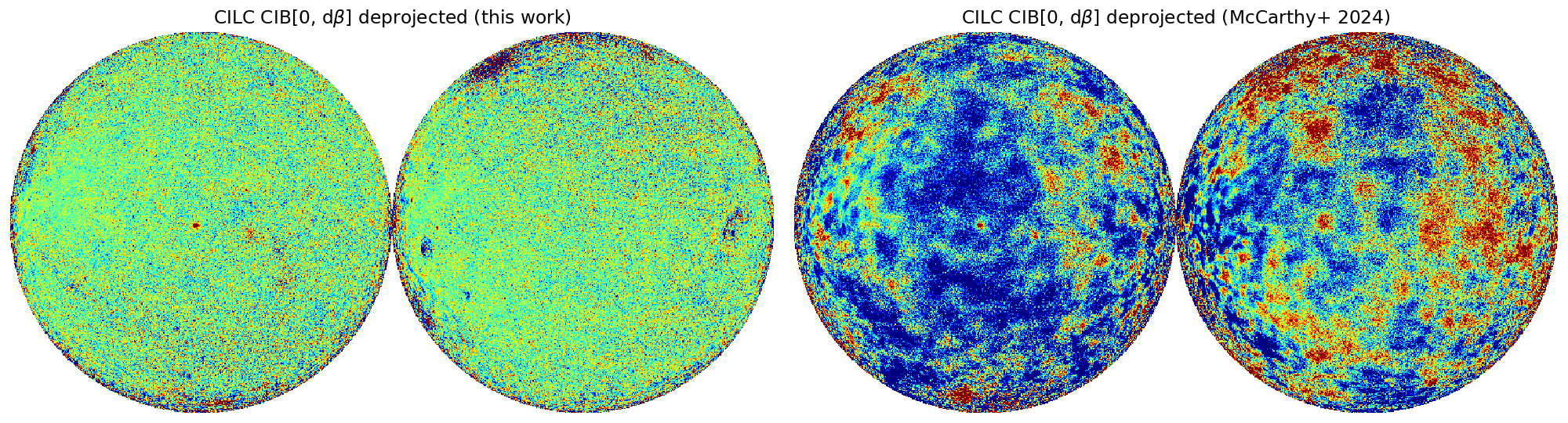}\\
    \includegraphics[width=0.92\textwidth]{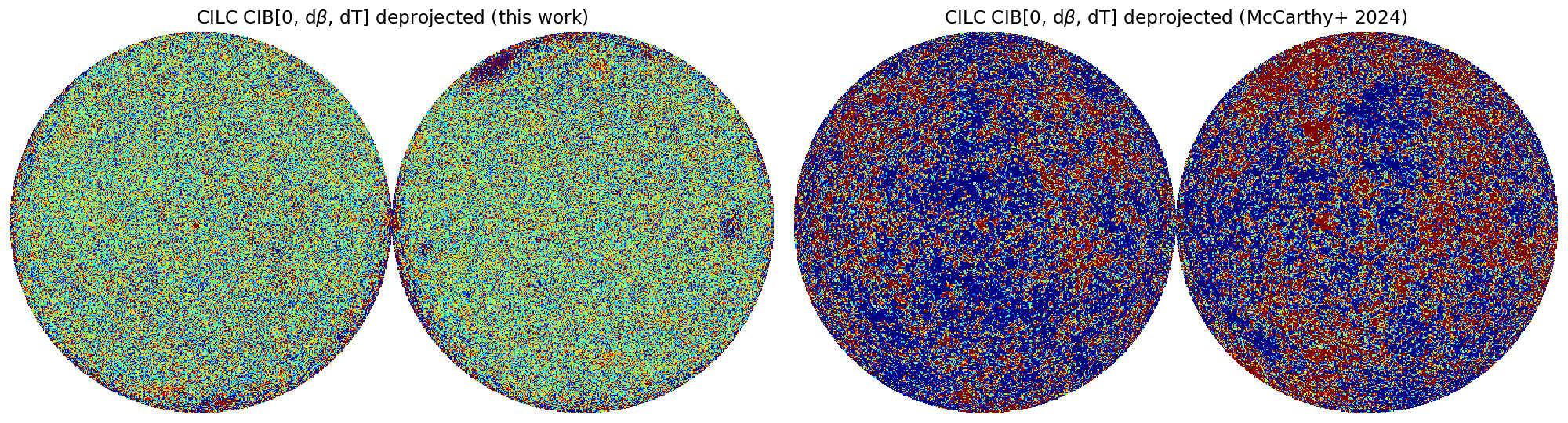}\\
    \includegraphics[width=0.92\textwidth]{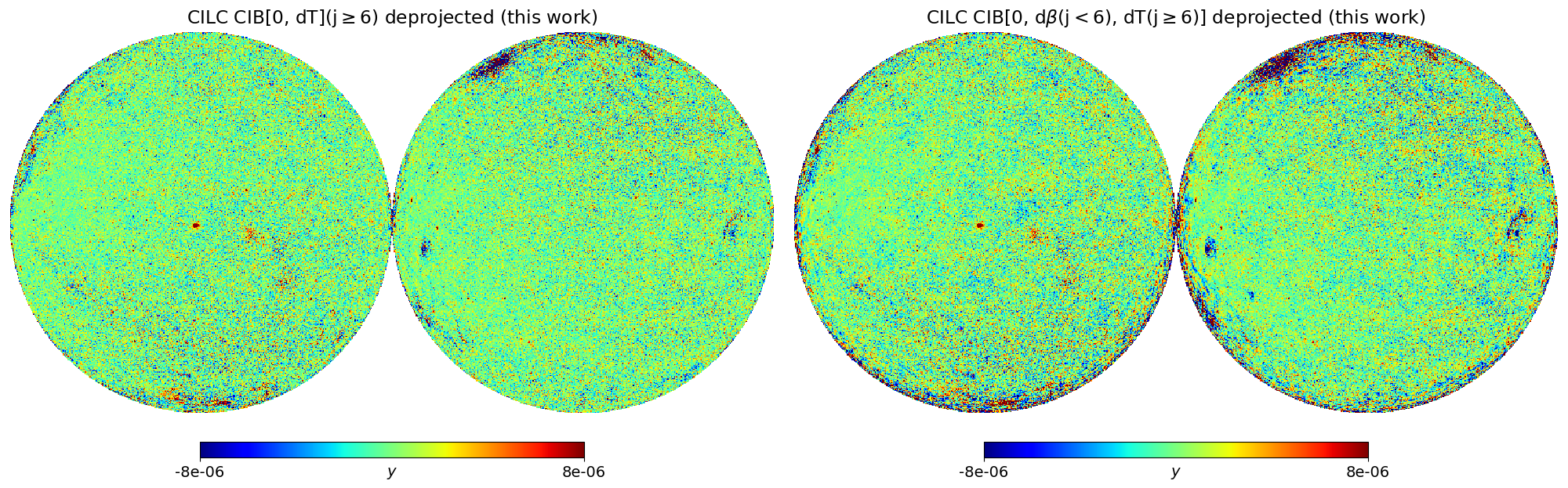}
    \caption{Comparison of PR4 NILC and CIB-deprojected CILC $y$-maps from our works (left; \cite{PR4NILCymap} and the present work) with those from McCarthy et al. (2024) \cite{NILC-McCarthyHill} (right) for comparable deprojection configurations. The last row shows our scale-dependent hybrid configuration (see Table~\ref{tab:cilc}).}
    \label{fig:cilc_maps}
\end{figure}

Recent CIB-deprojected $y$-maps based on a constrained ILC implementation were presented by \cite{NILC-McCarthyHill} using \pl\ PR4 data. Figure~\ref{fig:cilc_maps} provides a direct visual comparison between their maps and those constructed in our previous work \cite{PR4NILCymap} and in the present analysis, for closely matched deprojection configurations.\footnote{Maps from \cite{NILC-McCarthyHill} were obtained from \url{https://users.flatironinstitute.org/~fmccarthy/ymaps_PR4_McCH23/}.}

The first row shows the baseline PR4 NILC $y$-maps without CIB deprojection. Even at this level, noticeable differences are present: the NILC map from \cite{PR4NILCymap} (left) exhibits lower diffuse contamination and fewer compact residual features than the corresponding NILC map of \cite{NILC-McCarthyHill} (right). This indicates that differences in preprocessing, masking, needlet configuration, or implementation details of the NILC algorithm already affect the baseline reconstruction.

For the CIB-deprojected configurations (second to fourth rows), the differences become more pronounced. For comparable moment deprojections, the maps constructed in this work (left) display substantially reduced overall diffuse foreground contamination and noise. These differences are not limited to visual appearance but are confirmed quantitatively by the power-spectrum analysis shown below.

The methodological distinctions between the two analyses may help explain these outcomes. In the present work, the pivot parameters for the CIB SED ($\bar{T} = 26.1\,\mathrm{K}$ and $\bar{\beta} = 1.75$) are fixed to values derived from \pl\ measurements \cite{planck2013-pip56} and evaluated at the mean redshift of \pl\ clusters ($z \sim 0.2$), where the thermal SZ signal is most relevant, thereby defining a reference mean CIB SED and its higher-order moments. In contrast, \cite{NILC-McCarthyHill} determine the CIB spectral parameters through a best-fit procedure to the observed SED. While this approach offers flexibility, fixing the pivot parameters to physically motivated values tied to the cluster redshift appears, for PR4 data, to yield a reconstruction with lower variance for comparable CIB suppression.

In addition, rather than enforcing a fixed-order moment deprojection across all multipoles, we systematically explore combinations of specific moments and apply constraints selectively in angular scale. This leads to the hybrid configurations described in Section~\ref{sec:deproj_data} and Table~\ref{tab:cilc}, in which specific moments are deprojected only at high multipoles where CIB contamination dominates. These hybrid maps, shown in the last row of Figure~\ref{fig:cilc_maps}, are designed to minimize CIB-induced bias while limiting the variance penalty.

\begin{figure}[htbp]
    \centering
    \includegraphics[width=0.97\columnwidth]{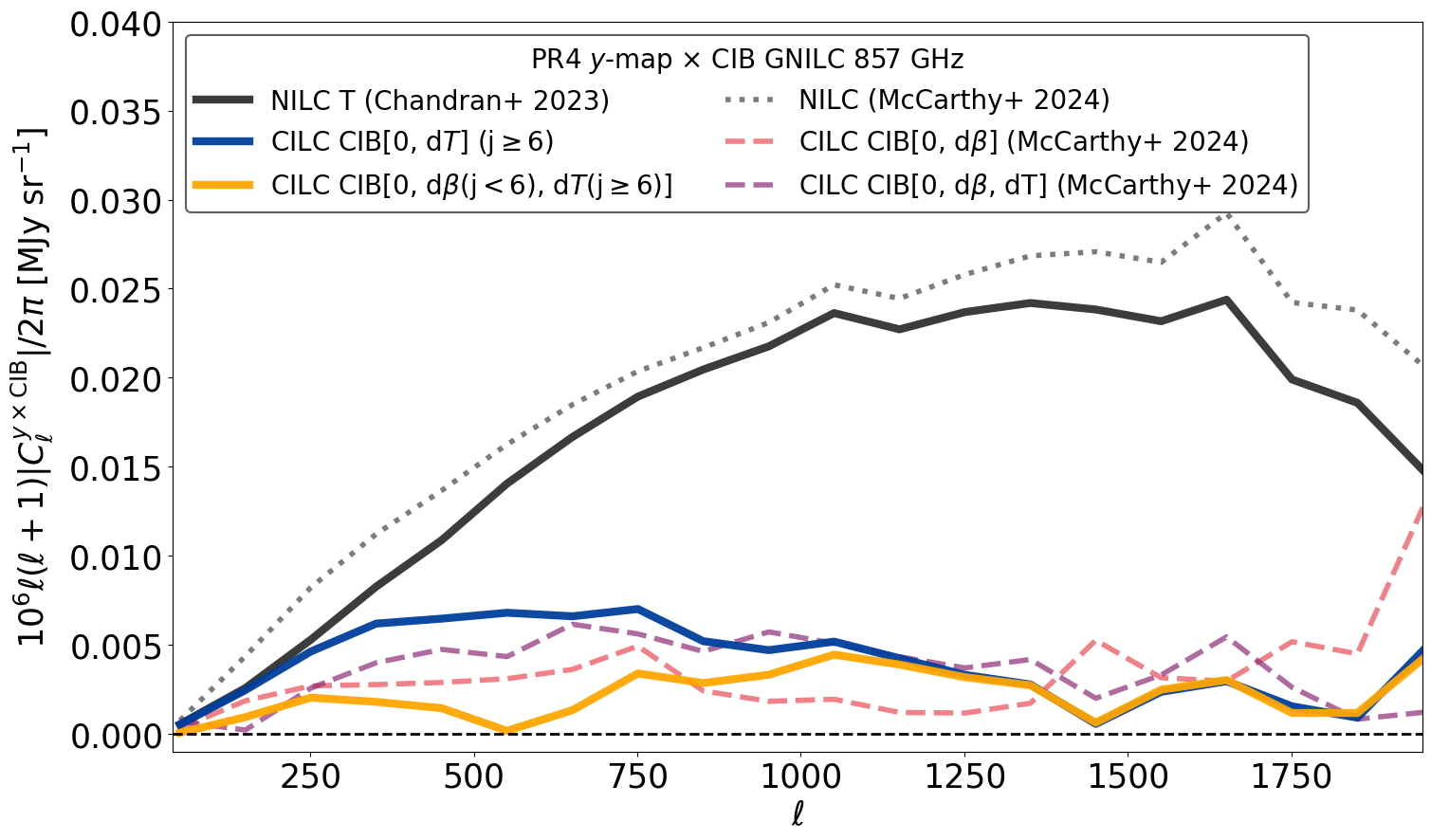}
    \includegraphics[width=0.97\columnwidth]{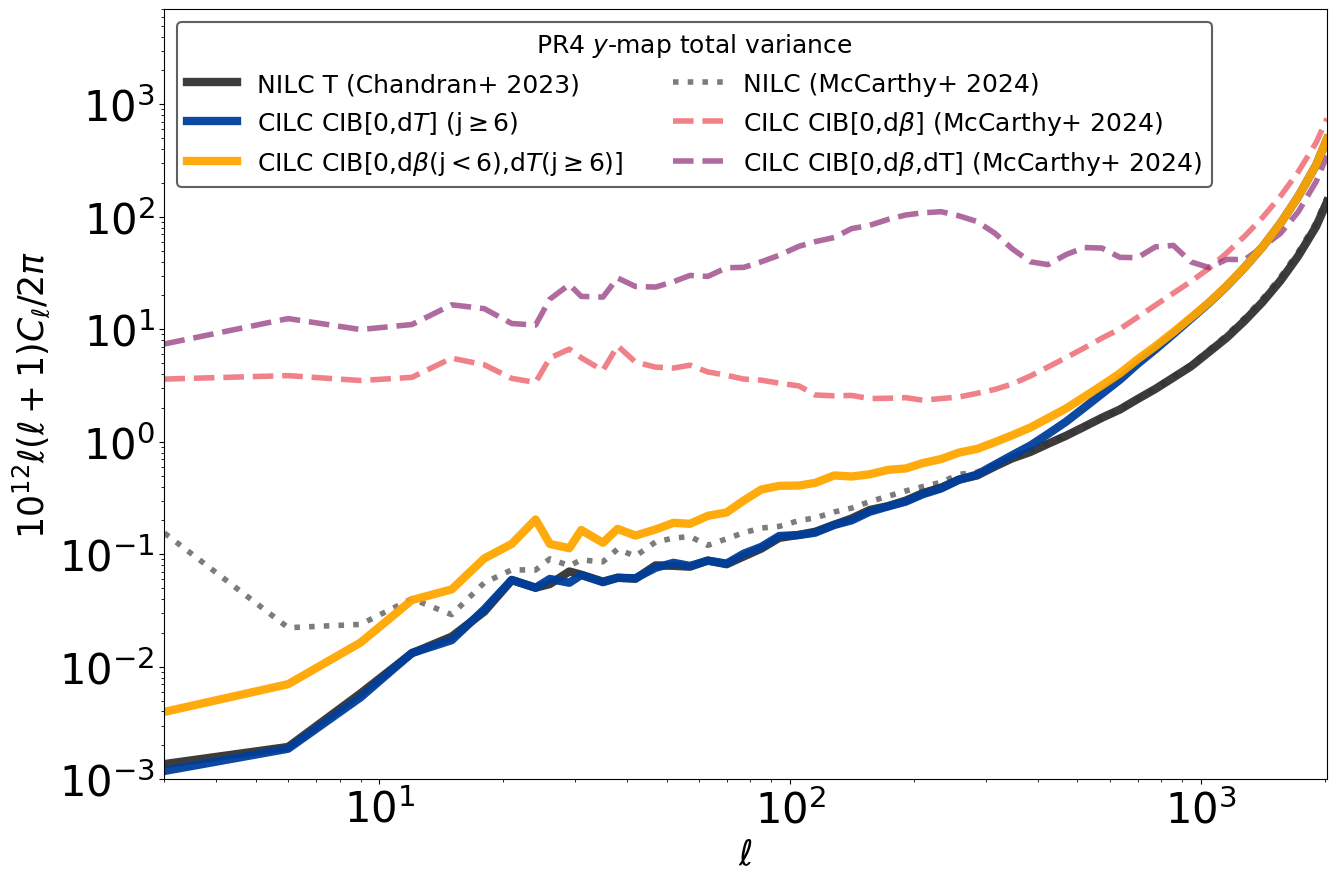}
    \caption{\emph{Top}: Cross-power spectra of PR4 $y$-maps with the GNILC CIB $857$\,GHz map over $50\,\%$ of the sky, comparing residual CIB contamination for maps from this work (solid lines) and \cite{NILC-McCarthyHill} (dashed/dotted lines). \emph{Bottom}: Corresponding auto-power spectra of the respective $y$-maps, illustrating differences in total variance for comparable deprojection configurations.}
     \label{fig:cib_deproj_cl_comparison}
\end{figure}

A quantitative comparison is presented in Figure~\ref{fig:cib_deproj_cl_comparison}. The top panel shows the cross-power spectra between each $y$-map and the GNILC CIB $857$\,GHz map, providing a quantitative measure of residual CIB contamination. The bottom panel shows the corresponding auto-power spectra of the respective $y$-maps, which reflect the total variance of the reconstructed maps.

For both the baseline NILC and the CILC configurations, the maps presented in this work exhibit systematically lower total auto-power across multipoles while achieving comparable or lower CIB cross-correlation levels. The $y$-maps from \cite{NILC-McCarthyHill} exhibit excessive power well above the expected thermal SZ signal across a broad range of multipoles (dashed lines), indicating that CIB deprojection in these cases typically induces a substantial noise and foreground penalty. This indicates that, for the PR4 data set, the selective and scale-dependent moment deprojection strategy adopted here leads to a more favourable noise--bias trade-off.

Overall, this comparison demonstrates that implementation choices within the CILC framework,  including pivot parameter selection, moment combinations, and scale-dependent constraints, have a substantial impact on the final $y$-map properties. For PR4, the strategy developed in this work yields lower-variance reconstructions while maintaining effective control of CIB contamination.

\end{document}